\documentclass[amsmath,twocolumn,superscriptaddress]{revtex4-1}
\usepackage{epsfig}
\usepackage{amssymb}
\usepackage{amsmath}
\usepackage{amsfonts}
\usepackage{braket}
\usepackage[T1]{fontenc}
\usepackage{bm}
\usepackage{graphicx}
\usepackage{xcolor}
\usepackage{verbatim}
\usepackage{soul}
\usepackage{tikz}
\usetikzlibrary{matrix,shapes,arrows,positioning,chains}
\usetikzlibrary{shapes.geometric}
\usetikzlibrary{shapes.arrows}
\usepackage{array}
\usepackage{enumitem}
\usepackage{color}
\usepackage{physics}
\usepackage[export]{adjustbox}
\usepackage[percent]{overpic}
\usepackage{filecontents}
\usepackage[normalem]{ulem} 
\usepackage{tcolorbox}
\usepackage{multirow}
\usepackage{hyperref}
\usepackage{lipsum}
\usepackage[caption=false]{subfig}
\hypersetup{
    colorlinks=true,
    linkcolor=blue,
    urlcolor=cyan,
    citecolor=blue,
    }
\usepackage{xfrac}

\begin{document}
\title{Deterministic Tensor Network Classifiers}

\author{L. Wright}
\affiliation{Department of Mathematics, King's College London, Strand, London WC2R 2LS, United Kingdom}

\author{F. Barratt}
\affiliation{Department of Physics, Amherst, Massachusetts, United States of America}

\author{J. Dborin}
\affiliation{London Centre for Nanotechnology, University College London, Gordon St., London, WC1H 0AH, United Kingdom}

\author{V. Wimalaweera}
\affiliation{London Centre for Nanotechnology,
  University College London, Gordon St., London, WC1H 0AH, United
  Kingdom}

\author{B. Coyle}
\affiliation{London Centre for Nanotechnology, University College London, Gordon St., London, WC1H 0AH, United Kingdom}

\author{A.~G. Green}
\affiliation{London Centre for Nanotechnology, University College London, Gordon St., London, WC1H 0AH, United Kingdom}
\affiliation{email: andrew.green@ucl.ac.uk}

\date{\today}
\begin{abstract}
We present tensor networks for feature extraction and refinement of classifier performance. These networks can be initialised deterministically and have the potential for implementation on near-term intermediate-scale quantum (NISQ) devices. Feature extraction proceeds through a direct combination and compression of images amplitude-encoded over just  $\log N_{\text{pixels}}$ qubits. Performance is refined using `Quantum Stacking', a deterministic method that can be applied to the predictions of any classifier regardless of structure, and implemented on NISQ devices using data re-uploading. These  procedures are applied to a tensor network encoding of data, and benchmarked against the 10 class MNIST and fashion MNIST datasets. Good training and test accuracy are achieved without any variational training. 
\end{abstract}
\maketitle

\vspace{0.2in}
\tableofcontents

\section{Introduction} 
\label{sec:introduction}
Classification using quantum circuits may be achieved by applying a unitary to data encoded as a qubit state and measuring a subset of the outputs\cite{nielsen_chuang,farhi2018classification,schuld2019quantum}.
This is easier said than done, however, and there are many subtleties to making this procedure workable and useful.
The choice of model (e.g. the specific data encoding, circuit ansatz and measurement strategy) is critical for training in order to avoid barren plateaux\cite{mcclean2018barren,holmes2021connecting,marrero2021entanglement,patti2021entanglement,wang2021noise} and for good initialisation\cite{dborin2021matrix} and refinement.
Moreover, noisy intermediate-scale quantum (NISQ) machines place additional constraints upon the process\cite{preskill2018quantum}, with limited numbers of qubits and finite quantum coherence. 
We show how tensor network methods permit deterministic initialisation of the model.
Specifically, we demonstrate this with two of the most well known tensor network decompositions via matrix product states (MPS)\cite{schollwock2011density,orus_review,Perez-Garcia2006} and tree tensor networks (TTN)\cite{shi_classical_2006}. The result is within the capabilities of current quantum devices.

Stoudenmire {\it et al}\cite{stoudenmire2016supervised,huggins2019towards} and Novikov\cite{novikov2016exponential} introduced the use of matrix product states for classical classification and found good performance on standard tasks.
These states, alongside with those produced by tree tensor networks, have a natural translation to quantum circuits\cite{schon2005sequential,banuls2008sequentially,wei2022sequential,hierarchical_quantum_classifiers}.
This ability to translate tensor networks between classical and quantum applications allows each to be used to their advantage.
The hope is that there may be quantum advantage in the speed with which the classification can be carried out on a quantum device\cite{huang2021quantum}, but also potentially in the training of tensor network based classifiers on quantum devices for use on classical devices. Conversely, as we demonstrate, classical manipulation of tensor network data encodings is an effective way to initialise the quantum classifier for further quantum training\cite{dborin2021matrix}. 

In this paper we give two insights into quantum machine learning with tensor networks.
{\it Deterministic feature extraction:}
We give an explicit method of tensor network encoding images by using sequential singular value decompositions.
A direct extension of this allows us to combine images in a given class to form a prototypical tensor network image state for that class. 
These are orthogonalised to form a single, unitary matrix product operator (MPO) or a tree tensor operator (TTO). 
This procedure amounts to a deterministic feature extraction and, crucially, contains a natural route for compression.
The MPO/TTO may be used directly to classify images and is a good starting point for further refinement. 
{\it Deterministic classifier initialisation:} We show how to improve classification by training on the output of potentially multiple feature extractors applied to image data --- a process known as stacking\cite{wolpert1992stacked}.
We achieve this classically by training a single-layer neural network on the output data and quantum mechanically using deterministic initialisation of a data re-uploading scheme\cite{perez2020data,schuld2021effect,caro2021encoding}.
Applying kernel methods to the output of classifiers leads to improved classification performance compared to using solely the classifier\cite{long2021properties}.
This motivates the use of supervised quantum machine learning on the classifier outputs, which can be thought of as a kernel method\cite{havlicek_supervised_2019,schuld2021supervised, jerbi2021quantum}, and may also be compared with alternative approaches to large scale image classification on quantum computers\cite{haug_large-scale_2021, li_quantum_2021, johri_nearest_2021, peters_machine_2021, marshall_high_2022}  

\section{Results}
\label{sec:Results}

Matrix product states (MPS) have enjoyed enormous success in their application to the classical simulation of quantum systems\cite{schollwock2011density,orus_review}.
This success rests on two facts:
i. MPS can be readily manipulated through sequential singular value decompositions into iso-metric form\cite{mps_classically_efficient,isometric_peps,haghshenas2019conversion} equivalent to the action of a unitary operator acting upon a reference state\cite{schollwock2011density,orus_review}. These manipulations are the bedrock of classical MPS algorithms for quantum simulation and their translation to NISQ devices\cite{smith2021crossing,barratt2021parallel,Lin2021real,foss2021entanglement};
ii. ground states of one-dimensional Hamiltonians can --- provably in some cases\cite{hastings2007area,hastings2007entropy} --- be captured by MPS of low Schmidt rank.
Quantum application of these methods potentially gains a quantum advantage, since the MPS encountered in time-evolution can be represented by shallow circuits\cite{Lin2021real} --- {\it even when their Schmidt rank is large} --- and manipulated exponentially quicker on quantum circuits than classically. 
Though utilised less frequently in classical simulation of quantum systems, tree tensor networks share advantageous features with MPS. Sequential singular value decompositions starting from the leaves of the tree and working towards the trunk yield an isometric form that can, similarly, be represented as a quantum circuit\cite{reyes2021multi,stoudenmire2018learning}. 

The algebraic advantages of MPS and TTN can be used directly in image classification. They allow images to be encoded as MPS/TTN on a small number of qubits. By translating classical image correlations into the entanglement structure of an encoded image state in this way, one may hope to benefit from some of the advantages of MPS found in quantum simulation.
Though this question of quantum advantage is beyond the scope of the present work.

\subsection{Tensor Network Encoding of Data}
\label{sec:Encoding}

An amplitude encoding of image data to $\log_2\left(N_{\text{pixels}}\right)$ can be rapidly converted to tensor network form through classical pre-processing.
In this form, the encoded data can be directly implemented on a quantum computer.
Moreover, this encoding provides a natural way to both compress and label image classes.
Compression allows key features of data to be learnt -- without compression it is hard to tell whether the image or noise are being fitted. In the following, for the purposes of clarity, we focus upon describing the method for the MPS and MPO version of the image encoding and classifier respectively. We describe the analogous methods for the tree tensor network in Appendix~\ref{app:TreeTensor}.

\begin{figure}[h]
\includegraphics[width = 1.0 \linewidth]{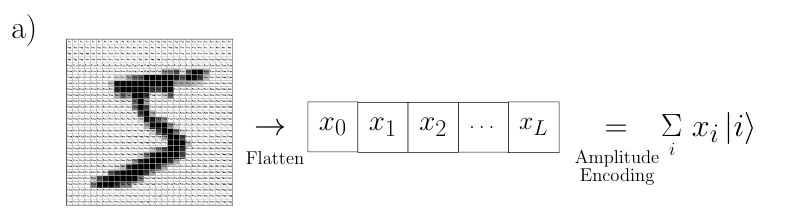}
\includegraphics[width = 1.0\linewidth]{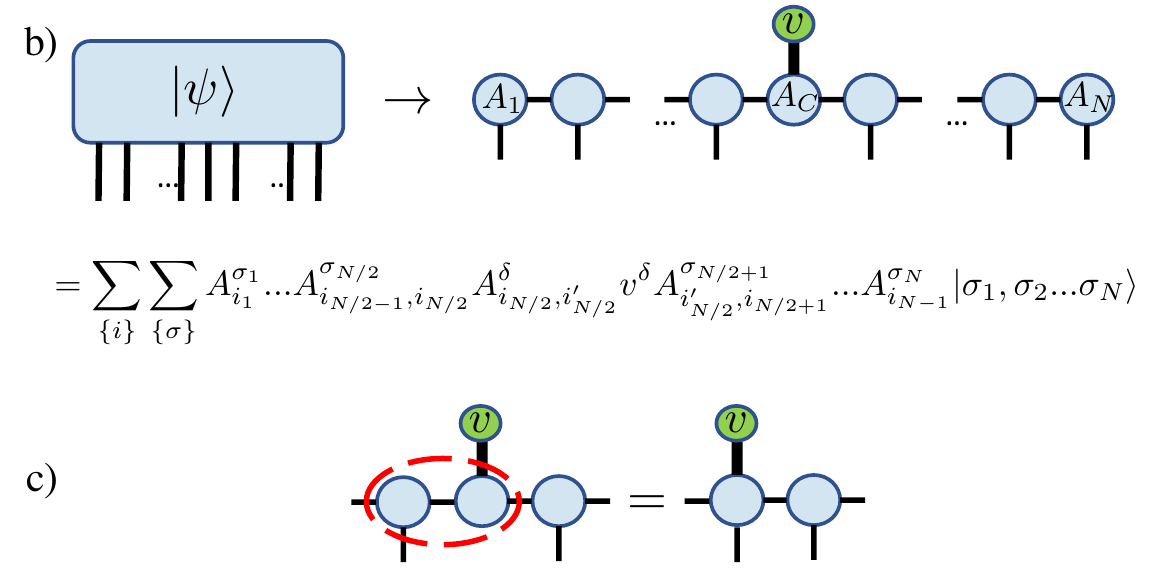}
\caption{{\bf Matrix Product State (MPS) encoding.} \newline
a) {\it Amplitude encoding:} A flattened vector of pixel intensities is used to produce a wavefunction encoding of the image. Pixels are labelled with bit-strings $i$ that provide basis vectors $| i \rangle$ for the amplitude encoding of a state $|\psi \rangle $ over $\log_2 N_{\hbox{\text{pixels}}}$ qubits. \newline
b) {\it MPS encoding:} A sequence of singular value decompositions\cite{mps_classically_efficient} from the left to the centre, and from the right to the centre of the state $|\psi \rangle$ converts this amplitude encoding to an MPS in mixed canonical form. 
To the left of the centre ($n=1 $ to $ n=N/2$), the resulting tensors are unitaries 
$ A^\sigma_{n \; ij} = U_{n\;  \sigma \otimes i,j} \in SU( 2^n)$. 
To the right of the centre ($n=N/2+1$ to $n=N$), the tensors are unitaries of the form 
$A^\sigma_{n\;ij} = U_{n\;  i, \sigma \otimes j} \in SU( 2^{N-n+1})$. 
Finally, the centre tensor is an isometry of the form 
$A_{c;ij} = U_{c;  i  \otimes j, \delta } v_{\delta}  \in SU( 2^{N/2})$, where
$v_\delta$ is a set of amplitudes for some $N$-spin reference state which we conventionally take to be $|0 \rangle^{\otimes N}$. In these tensors, $\delta$ and $\sigma$ index the physical reference space and output space, respectively. $i$ and $j$ index the auxiliary space. The Schmidt rank of the auxiliary space between sites $n$ and $n+1$ is $D_n=2^n$ up to $n=N/2$ and $D_n=2^{N-n}$ beyond. The procedure is analogous for the tree tensor decomposition, see Appendix~\ref{app:TreeTensor}. \newline
c) {\it Absorbing Centre Tensor:} the centre tensor may be absorbed into the tensor to its left (or right) without loss of generality.}
\label{fig:MPSEncoding}
\end{figure}

\begin{figure*}
    \captionsetup{justification=raggedright}
    \includegraphics[width = 0.8\textwidth]{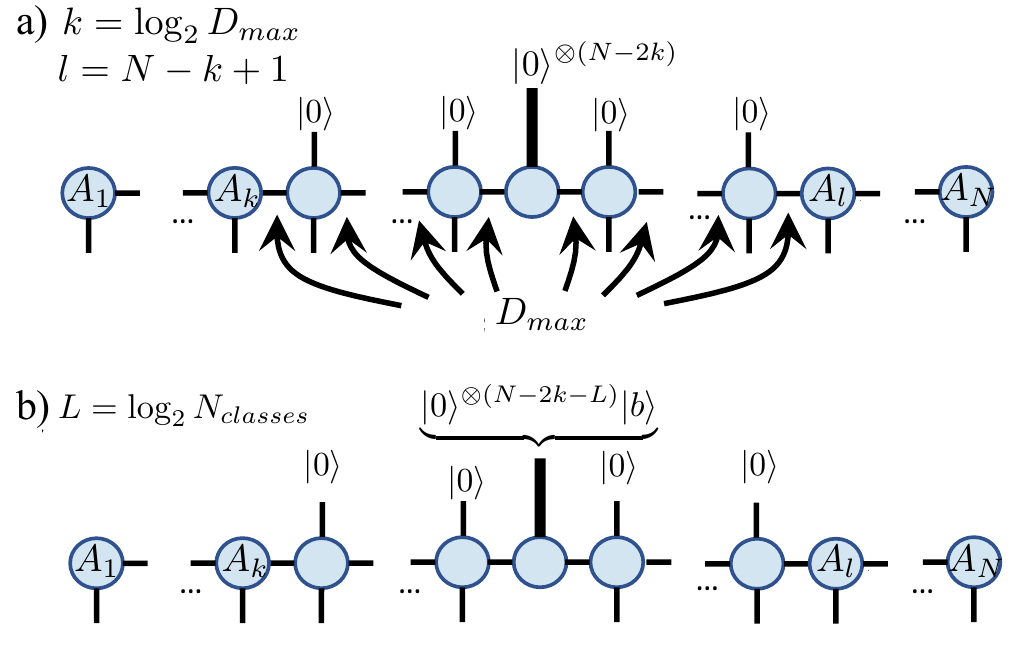}
    \caption{
    {\bf Compression and labelling of images.} \newline
    a) {\it Compression of MPS:} The amplitude encoding can be compressed by truncating a sequence of singular value decompositions on the auxiliary space from right to left\cite{schollwock2011density,orus_review,Perez-Garcia2006}  to some maximum value $D_{max}$ .This implies a corresponding reshaping of the tensors and the reference state as shown.\newline
    b) {\it Labelling:} The use of different reference states provides a straightforward way to label wavefunction encodings of images in different classes. The final $\log_2 N_{\text{classes}}$ reference states are used for a binary label, the first $\log_2 (N_{\text{pixels}}/N_{\text{classes}})$ remaining as $|0 \rangle$. It is convenient to reshape the tensors as indicated so that the label is entirely on the final tensor. This reshaping facilitates multi-state classification on a quantum circuit with a single unitary.}
    \label{fig:MPSCompression}
\end{figure*}

{\it The procedure for MPS encoding an image} is illustrated in Fig.~\ref{fig:MPSEncoding}. First, a two-dimensional image is flattened into a one-dimensional vector of pixel intensities.
We choose to flatten the vector in row-major order.
An amplitude encoding maps the pixel intensities to the amplitude of bitstrings of length $N_{\text{qubits}}= \log_2\left(N_{\text{pixels}}\right)$.
This superposition of bitstrings can then be encoded as an MPS using a sequence of singular value decompositions\cite{mps_classically_efficient}.
As illustrated in Fig.~\ref{fig:MPSCompression}, such an encoding has an in-built compression scheme determined by the Schmidt rank or bond order of the MPS.
Note, we use the terms Schmidt rank, bond order, and bond dimension interchangeably. 
An uncompressed encoding is obtained when the bond order between the $n^{th}$ and $n+1^{th}$ qubit increases as $D_n=2^n$ up to $D_{\max}= D_{N_{\text{qubits}}/2}= 2^{N_{\text{qubits}}/2}=\sqrt{ N_{\text{pixels}}}$ and then decreases as $D_n=2^{N_{\text{qubits}}-n}$.
Truncating the bond order to some maximum value gives a compression without averaging over pixels and is akin to a local principal component analysis of the image.

The pixel intensities of the encoded images can be restored by computing the overlap with the bitstring states.
Moreover, the mapping from image correlations to the entanglement structure of the MPS encoding may reveal useful properties of the encoded state.
Here we have focused upon an amplitude encoding to the minimum number of qubits.
Alternative mappings to more qubits --- and so to a subset of possible bitstrings --- may preserve features such as locality and produce encoded states that can be manipulated with quantum advantage. 
As mentioned above, we demonstrate the mapping from bitstring superposition to the alternative example of a tree tensor network in Appendix~\ref{app:TreeTensor}. One could also consider mapping to other structures beside these two, for example via multi-scale renormalisation ansatz (MERA) states\cite{mera_vidal,hierarchical_quantum_classifiers,cong2019quantum}. Such alternative decompositions capture correlations within the data differently which might better suit specific tasks.

\begin{figure}[h]
\centering
\includegraphics[width=0.8\linewidth]{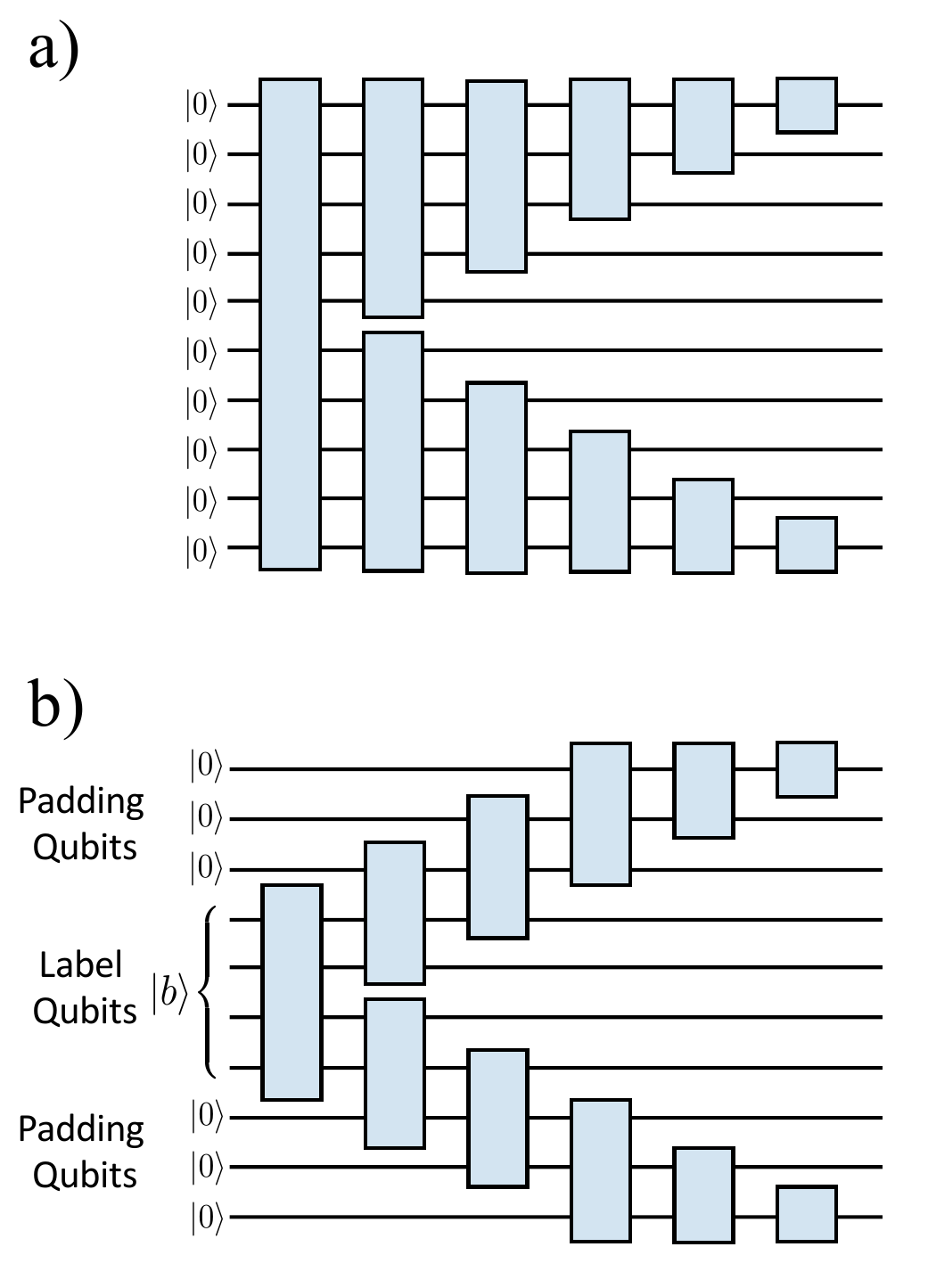}
\caption{{\bf Circuit Representation of MPS-encoded images.} MPS in orthogonal form can be translated directly to quantum circuits. a) shows the translation of the full quantum state $|\psi \rangle$ and is equivalent to Fig.~\ref{fig:MPSEncoding}b. b) shows a truncation to bond order $D_{\max}=4$ and how reference qubits are used to label the classes. 
}
\label{fig:CircuitRepresentation}
\end{figure}

{\it MPS-encoded images can be translated directly to quantum circuits}. As indicated in Fig.~\ref{fig:MPSCompression}, the singular value decomposition yields isometric MPS tensors.
They can be translated to a quantum circuit as shown in Fig.~\ref{fig:CircuitRepresentation} a).
Quantum circuit MPS have been used previously in quantum simulation\cite{smith2021crossing,barratt2021parallel,Lin2021real,foss2021entanglement} and for initialisation of quantum circuits\cite{dborin2021matrix}, and one may also construct quantum circuits from tree tensor networks, as we discuss in Appendix~\ref{app:TreeTensor}.
We will return to the circuit representation later when we discuss refinement of image classification by training on output data.

\subsection{Tensor Network Feature Extraction}
Here, we offer a deterministic scheme to initialise our tensor network feature extractor through a sequential addition of encoded data.
The addition is first applied batch-wise on data of the same class, ensuring compression after each batch.
Resultant sum states for each class are then encoded with their corresponding bitstring label states, added together and orthogonalised to produce our feature extractor. For clarity, again we detail the construction of feature extractor using the MPS decomposition, see Appendix~\ref{app:TreeTensor} for the TTN version.

\subsubsection{MPS Encoding of Sum States}
\label{sec:SumState}

We deterministically construct a prototypical state $\ket{\Sigma_l}$ for a given class $l$ as an unweighted sum of all images in the class. The underlying assumption is that the image states are clustered in Hilbert space. By interpreting the resulting MPS as the action of an MPO on a reference state, labels may be added for the different classes by using different reference states. MPS-encoded images can be classified by finding the prototypical state with which they have maximum overlap.

{\it Constructing the Sum State}: 
For a set of three MPS characterised, for example, by tensors $\{ A_1, A_2,...\}$, $\{ B_1, B_2,...\}$ and $\{ C_1, C_2,...\}$ the MPS representation of the sum state is formed by first combining the tensors into one larger tensor as follows:
\\
\noindent
i. First site:
\begin{eqnarray*}
& &
{\cal A}^{\sigma_1}_{1\; a_1\otimes b_1 \otimes c_1}
\\
&=&
( A^{\sigma_1}_{1\;1},A^{\sigma_1}_{1\;1},...A^{\sigma_1}_{1\;D_1},B^{\sigma_1}_{1\;1},...B^{\sigma_1}_{1\;D_1},C^{\sigma_1}_{1\;1},...C^{\sigma_1}_{1\;D_1})
\\
&=&
(\underline{A}^{\sigma_1}_{1},\underline{B}^{\sigma_1}_{1},\underline{C}^{\sigma_1}_{1})
\end{eqnarray*}
\noindent
ii. General site:
$$
{\cal A}^{\sigma_n}_n
=
\left(
\begin{array}{ccc}
\underline{\underline{A}}^{\sigma_n}_{n} & 0 & 0\\
0 & \underline{\underline{B}}^{\sigma_n}_{n}  & 0 \\
0 & 0 & \underline{\underline{C}}^{\sigma_n}_{n} 
\end{array}
\right)
$$
\noindent
 iii. Centre Site:
$$
{\cal A}^{ \delta}_C
=
\left(
\begin{array}{ccc}
\underline{\underline{A}}^{\delta}_{C} & 0 & 0\\
0 & \underline{\underline{B}}^{\delta}_{C}  & 0 \\
0 & 0 & \underline{\underline{C}}^{\delta}_{C} 
\end{array}
\right)
\left(
\begin{array}{ccc}
v^{\bm{\delta}}&0 &0  \\
0& v^{ \bm{\delta}} &0 \\
0& 0& v^{\bm{\delta}} 
\end{array}
\right),
$$

\noindent
 iv. Last site:
$$
{\cal A}^{\sigma_N}_N
=
\left(
\begin{array}{ccc}
\underline{A}^{\sigma_N}_{N} \\
\underline{B}^{\sigma_N}_{N}  \\
\underline{C}^{\sigma_N}_{N} 
\end{array}
\right),
$$
where $\underline{M}$ is a vector and $\underline{\underline{M}}$ is a matrix. Auxiliary-space tensor indices have been suppressed in the second, third and fourth cases for clarity.  The reference state amplitudes $v^{\bf \delta}$ are the same for each image if they are in the same class. Conventionally, we choose a reference state $|0 \rangle^{\otimes N}$.

This block diagonal tensor is compressed back down to the desired bond order using a sequence of singular value decompositions as described in Fig.~\ref{fig:MPSCompression} a)-c). 
Note that when the MPS bond order is high enough to cover the Hilbert space ($D=32$ in the case of MNIST and fashion MNIST), no compression is implied in this procedure. 

\begin{figure}[h]
\centering
\includegraphics[width=\linewidth]{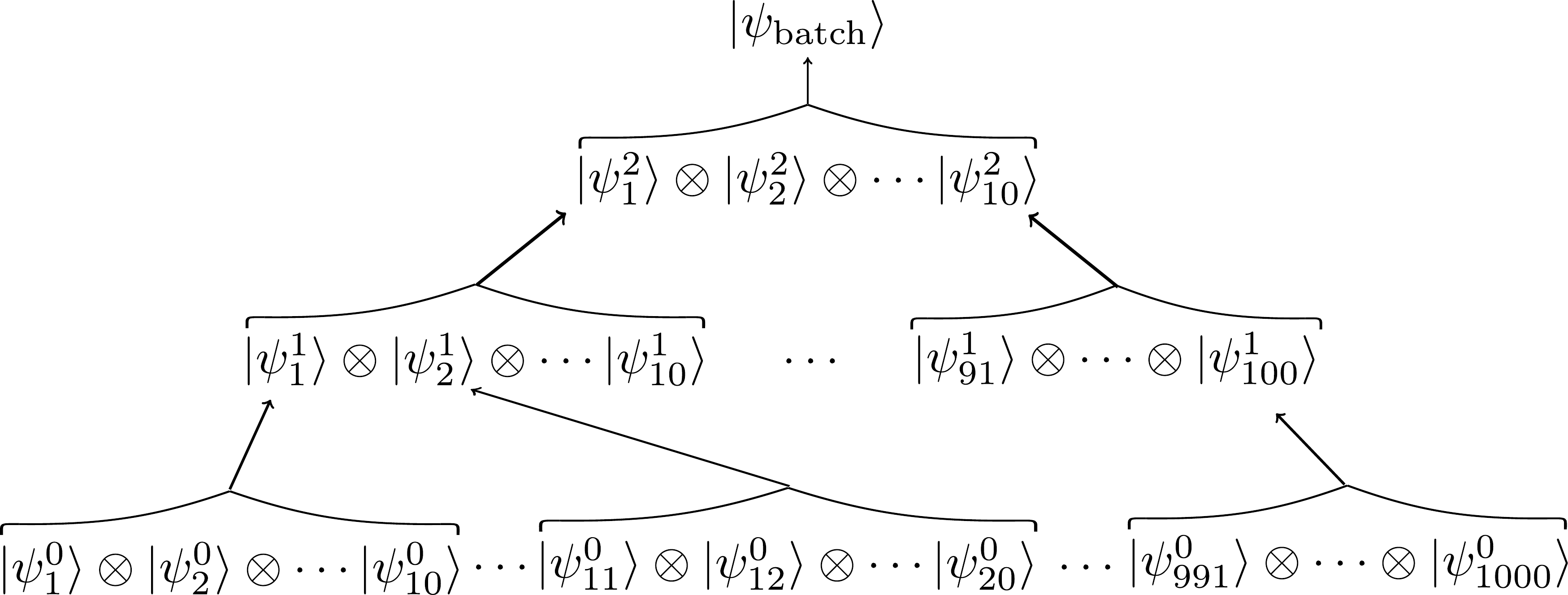}
\caption{{\bf Batching of MPS-encoded images:} 
For large training sets it may not be possible to construct the full sum state before re-arranging it as an MPS.
In this case, we proceed by first finding the sum state over smaller batches and progressively combining and compressing these batches.
When the maximum bond order of each batch is large enough ($D_{\text{batch}} \ge D_{max} = 32$ in the case of MNIST)
 the representation of the sum state is exact and this procedure gives the same result as forming the sum-state MPS in a single step.
When $D_{\text{batch}} < D_{\max}$, the final result gives a compressed approximation to the sum state. This process yields a slightly different state for different batchings of the images. Here we show a training set consisting of $1000$ images progressively combined in batches of $10$.}
\label{fig:Batching}
\end{figure}

{\it Batch construction of sum states}:  In practice the block diagonal combined MPS may be too large to manipulate. In this case the sum state can be constructed by sequentially combining and compressing batches of MPS-encoded image states as shown schematically in Fig.~\ref{fig:Batching}.
When the maximum bond order of each batch, $D_{\text{batch}}$, equals or exceeds 
$D_{max} =2^{N_{qubits}/2}=\sqrt{N_{\text{pixels}}}$ the sum state is obtained without approximation.
When $D_{\text{batch}} < \sqrt{N_{\text{final}}}$, this batching construction involves compression at each step and the final result gives an approximation to the sum state. This compression makes overfitting less likely.
The truncation of singular values obtained through the singular value decomposition acts as a regulariser--- disregarding the information deemed to be least important within the data. The frequent projections back down to the set of MPS states 
are reminiscent of various schemes for approximating quantum dynamics using MPS\cite{stoudenmire2016supervised,huggins2019towards}, and in effect comprise a local principal component analysis. 

{\it Labelling different classes:}  As illustrated in Fig. \ref{fig:MPSCompression} a) and discussed in Sec. \ref{sec:Encoding}, an MPS in canonical form may be interpreted as the action of a matrix product operator (MPO) on a reference state.
This allows direct translation to quantum circuits  as shown in Fig. \ref{fig:CircuitRepresentation}.
Moreover, it provides a convenient way to label different image classes using different bitstring reference states for each class. 

The bitstring labels, ${\bf b}_l $ can be used to construct an orthonormal set of reference states $\ket{{\bf b}_l}$ on the last 
$N_{\text{label}}= |\left(  \log_2 ( N_{\text{classes}} ) \right)|$ 
qubits.
The remaining qubits --- which we refer to as padding qubits --- are put in the state $\ket{p}=| 0 \rangle^{\otimes N_{\text{pad}}}$. 
For example, class $9$ is encoded into $| 0 \rangle^{\otimes N_{\text{pad}}} \otimes \ket{1001}$.

Contracting the bitstring with the MPO $U_l$ returns corresponding sum state,
	$$\ket{\Sigma_l} = U_l \ket{p} \ket{{\bf b}_l}.$$
Note that the unitaries, $U_l$, representing the sum states are different for each class label $l$.
Presently, we show how to construct a single unitary from which representative states for all classes can be constructed.

\subsubsection{Classification with Sum States}
\label{sec:Classification with Sum State}

The set of sum states $\ket{\Sigma_l}$ can be used to classify MPS-encoded images.
{\it In classical implementations}, optimising the overlap of a test image with this sum-state provides a simple way to classify. 
For an MPS-encoded image, $\ket{image}$, and padding, $\ket{p}=\ket{0}^{\otimes N_{\text{pad}}}$, the image is classified as class $k$, where
\begin{equation}
k = \max_l |\bra{image} U_l \ket{p}\ket{{\bf b}_l}|^2
\label{eq:PostSelectClassify}
\end{equation}
Classically, the expectations are evaluated by explicit contractions of the tensor indices.
On a quantum computer, the overlap requires post-selection on the state $\ket{p}\ket{{\bf b}_l}$ (which is exponentially costly) or else a swap test between $U_l  \ket{image}$ and $\ket{p}\ket{{\bf b}_l}$.

{\it For implementation on a quantum computer} it is preferable to simply measure the Pauli $Z$ operator on the label qubits and use the resulting output to assign a label.
This corresponds to the assignment
\begin{equation}
    k = \max_l \sum_p  |\bra{image} U_l \ket{p}\ket{{\bf b}_l}|^2,
\label{eq:TraceOutClassify}
\end{equation}
where a sum or trace over all paddings is taken.
Tracing over paddings is potentially exponentially quicker than post-selecting paddings on a quantum device, but we find that it performs worse than the post-selection of Eq.~(\ref{eq:PostSelectClassify}).

{\it Comments on the sum state classifier:}
In principle, one need not construct the sum state $|\Sigma_l\rangle$ in order to compute its overlap with a given test image. The sum of the overlaps with each test image gives the same result.
An explicit construction of the sum state reduces the number of overlaps with the test data from $\mathcal{O}(N_{\text{train}} N_{\text{test}})$ to $\mathcal{O}(N_{\text{test}})$. In addition, the explicit sum state has the advantage of ease of implementation on a quantum computer. 

Recently, Vidal {\it et al} showed that the sum state is not optimal for classification of product state encoded data\cite{martyn2020entanglement}. It is not evident that this result applies to an amplitude encoding, where classical correlations of individual images have been encoded in the entanglement structure. A sum state constructed from product state encodings has different entanglement structure than any individual state. In contrast, a sum state constructed from amplitude encodings has a similar entanglement structure. 

So far we have constructed different sum states $\ket{\Sigma_l}$ for each class $l$, corresponding to the action of different unitaries $U_l$ on the reference states $\ket{p} \ket{{\bf b}_l}$. Next, we combine these into a single unitary that allows direct multi-state classification on a single quantum circuit.

\subsubsection{Orthogonalising Sum States }
\label{sec:Orthogonalising}

The MPS $\ket{ \Sigma_l }$ and the MPOs, $U_l$,  that generate them can be orthogonalised and combined to form a single matrix product operator $U$.
The construction follows much the same steps as described in Sec \ref{sec:SumState}, with the tensors $A$, $B$ and $C$ now corresponding to the MPS representation of sum states for different classes.
The treatment of the central tensor is different since these carry the label index, ${\bm \delta}$.
The combined tensor for the central site then becomes
\begin{eqnarray*}
{\cal A}^{\delta}_C
=
\left(
\begin{array}{ccc}
\underline{\underline{A}}^{\bm{\delta}}_{C}& 0 & 0 \\
0&  \underline{\underline{B}}^{\bm{\delta}}_{C}&0  \\
0& 0& \underline{\underline{C}}^{\bm{\delta}}_{C} 
\end{array}
\right)
\left(
\begin{array}{ccc}
v^{\bm{\delta}}_{A}&0 &0  \\
0& v^{ \bm{\delta}}_{B} &0 \\
0& 0& v^{\bm{\delta}}_{C} 
\end{array}
\right),
\end{eqnarray*}
where as before auxiliary-space tensor indices have been suppressed for clarity.
The reference state vectors $v_A^{\bm{\delta}}$, $v_B^{\bm{\delta}}$ and $v_C^{\bm{\delta}}$ are taken to be orthogonal bitstrings for each class. The central tensor then has non-zero elements in different parts of the reference space for different classes.  
Up to the central site, the singular value compression proceeds as before.
For sufficiently large bond order in the MPS ($D_{\text{batch}} \ge D_{\max} = 32$ in the case of $28\times 28$ image data, such as MNIST) no information is lost in this process. 
Because of the additional reference-space structure in the central tensor, it cannot be converted to a unitary without loss of information (unless the sum states $\{\ket{\Sigma_l } \}$ form an orthonormal set).
We achieve a unitary form by using a polar decomposition --- a singular value decomposition where the singular values are replaced with the identity, giving the best unitary approximation to the central tensor.
The resultant MPO, viewed generatively through its action on the reference bitstrings, produces an orthonormal set of states.
It can be represented as a single unitary for all classes in a circuit implementation. 
The classification process assigns a class-label to a test image according to 
\begin{eqnarray}
    k &=& \max_l |\bra{image} U \ket{p}\ket{{\bf b}_l}|^2
    \nonumber\\
    & & {or}
    \nonumber\\
    k &=& \max_l \sum_p|\bra{image} U \ket{p}\ket{{\bf b}_l}|^2,
\label{eq:CombinedClassifier}
\end{eqnarray}
in the classical and quantum implementations respectively. The unitaries $U_l$ in Eqs. (\ref{eq:PostSelectClassify}) and (\ref{eq:TraceOutClassify}) have been replaced by the single unitary $U$.

{\it Comments:} 
In principle, the combined classifier can be constructed from the original image data in one go without the intermediate step of producing the sum states.
The resulting block diagonal MPO is prohibitively large and a batching procedure as described in Sec.~\ref{sec:SumState} and Fig.~\ref{sec:SumState} is required.
These batches need not contain images drawn only from the same class. For example, each batch could contain an image for every class or be completely random. We find, however, that forming the sum states for each class first and then combining them leads to better performance. 

\begin{figure*}[ht!]\hspace{-1cm}
    \centering
    \includegraphics[width = \linewidth]{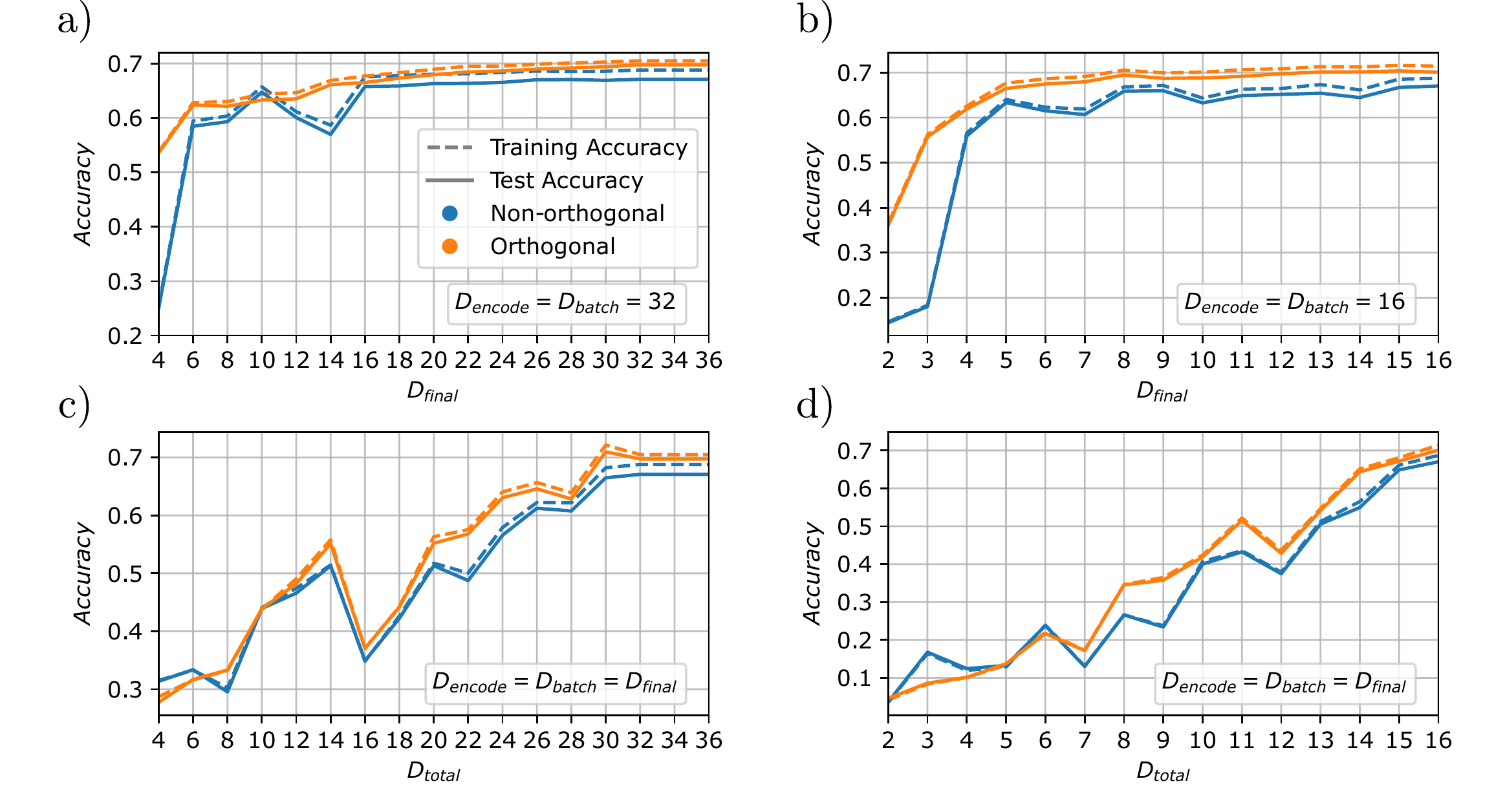}
\caption{ 
\textbf{Classification Accuracy and Bond Order for Fashion MNIST.}\\
Training and test accuracy of the tensor network classifier for different bond orders constructed through the deterministic batching procedure for the fashion MNIST dataset using the (a, c) MPO and (b, d) TTO decomposition. In all cases, $60000$ training images were used to construct the classifier, whilst all $10000$ test images were used to evaluate performance.
\\
{\it Top row} shows the accuracy of the a) MPO and b) TTO classifiers as a function of $D_{\text{final}}$ with $D_{\text{encode}}$ and $D_{\text{batch}}$ set to be maximal (where maximal bond dimension is $32$ for the MPO, and $16$ for the TTO, see Appendix~\ref{app:TreeTensor} for a discussion).
\\
{\it Bottom row} shows the test accuracy of the c) MPO and d) TTO classifiers with $D_{\text{encode}}$ set equal to $D_{\text{batch}}$ and $D_{\text{final}}$ with all three varied simultaneously. We also show classifier performance with and without orthogonalisation (polar decomposition on the final MPO tensor, or trunk tensor of TTO) which is required for implementation on a quantum device.
}
\label{fig:fashion_accuracy_vs_d_total}
\end{figure*}

{\it Results:} The performance of deterministic initialisations of the MPO and TTO classifiers are shown in Fig.~\ref{fig:fashion_accuracy_vs_d_total}. The ultimate performance depends upon the bond order $D_{\text{final}}$. The best performance attained for the Fashion MNIST dataset with our deterministic initialisation scheme using an MPS/MPO decomposition is $69.75\%$ test accuracy and is found for $D_{\text{final}} \gtrsim 20$. For the TTN/TTO decomposition, the highest achieved test accuracy is $70.1\%$, with a maximal bond dimension of $16$ (see Appendix~\ref{app:TreeTensor}).

Whether this performance is found after a particular deterministic initialisation depends also upon the bond order of the encoded image, $D_{\text{encode}}$, and the bond order used in the batching process $D_{\text{batch}}$. Both must be sufficiently large in order to saturate the potential performance. The test accuracy of an MPO is therefore a function of three parameters, $D_{\text{final}}$, $D_{\text{batch}}$, and $D_{\text{encode}}$. We observe in Appendix~\ref{app:tuning_feature_extraction} that performance is least sensitive to $D_{\text{encode}}$, and value as low as $D_{\text{encode}} = 5$ (for the MPO version) does not degrade the final accuracy provided $D_{\text{final}}$ and $D_{\text{batch}}$ are sufficiently large. The fact that maximum performance is obtained for $D_{\text{final}} \approx 20$, which is less than the bond order required to arbitrarily locate a point in Hilbert space, points to a redundancy in the data --- or alternatively to a limit in the number of identifiable features. Of particular note is that to obtain this performance at $D_{\text{final}}=20$ the bond order used in the batching process $D_{\text{batch}}$ must be larger than $20$. This presumably avoids compressing away important information earlier in the training process and is akin to the failure of other computational processes when the bond order is limited\cite{crowley2014quantum,barratt2021dissipative}.

\subsection{Classifier Refinement via Stacking}
\label{sec:ClassificationRefinement}

Once initialised, classification performance can be improved by updating either the elements of the feature extractor itself, or its outputs.
The former can lead to greater expressibility due to the increased size of feature space, however it can also be difficult to train\cite{barratt2022improvements}, require specific pre-processing of the data\cite{dilip_data_2022} and each individual structure of feature extractor may require bespoke optimisation strategies. 
Therefore we focus on the latter case using a procedure known as stacking --- the process of training an independent classifier on the outputs of potentially multiple feature extractors. Since the dimension of the label space is significantly smaller than the feature space of the data, fewer parameters require training compared to training the classifier elements explicitly.

\subsubsection{Classical and Quantum Stacking}
We begin by discussing the classical and quantum implementation of the stacking procedure and then show how the efficiency of the latter can be improved in the next sub-section. 

\vspace{0.1in}
{\it Classical Training on Classifier Outputs:}
As an example of classical stacking we train a neural network on the outputs of the feature extractor. 
In order to generate a data set for stacking, we act on each training data image with the deterministically initialised $U$.
The resulting state is overlapped with each padded bitstring $\ket{0}^{\otimes N} \otimes \ket{{\bf b}_l}$ to give a vector of bitstring amplitudes for each image:
\begin{equation}
v_l = |\bra{image} U \ket{0}^{\otimes N} \otimes \ket{{\bf b}_l}|^2.
\label{eq:classicalvector}
\end{equation}
Training a neural network on these output vectors using stochastic gradient descent allows a non-linear transformation of the predictions.
We use a single fully-connected layer, including bias terms, followed by sigmoidal activation function.
The total number of parameters used is 170 ensuring that the data is not overfitted. The results of this refinement are shown in Table \ref{tab:RefinementResults} and indicate a $6\%$ improvement.

\begin{figure}[h]
\hspace{-8cm}\raisebox{0pt}{a)}
\\
\includegraphics[width=0.95\linewidth]{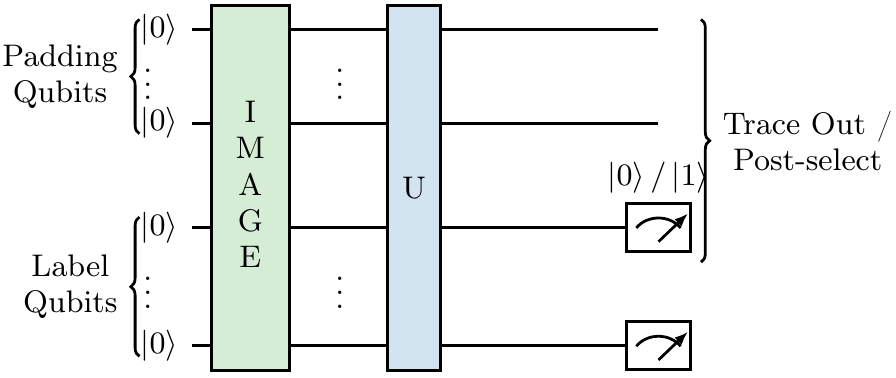}
\\
\hspace{-8cm}\raisebox{0pt}{b)}
\\
\includegraphics[width=\linewidth]{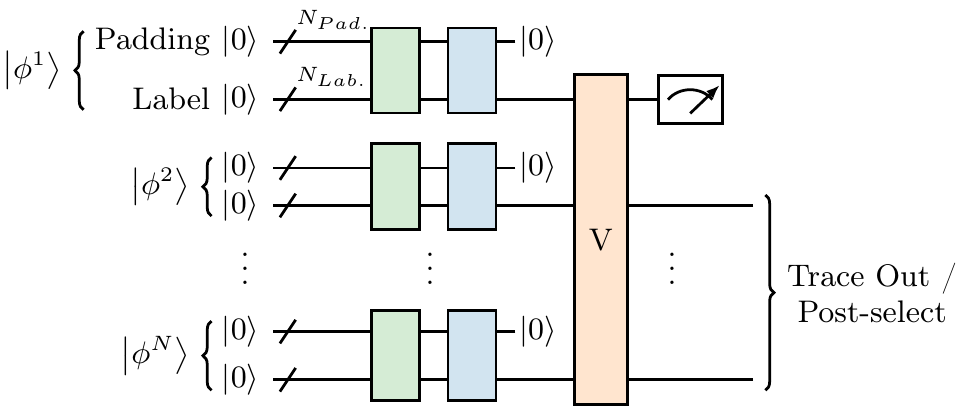}
\caption{{\bf MPO Classification and its Refinement} 
a) {\it Basic classification} of images using the orthogonalised MPO obtained from singular value decomposition of the training data.
The padding qubits may be post-selected to be $|0 \rangle$ or traced over.
The former is convenient in classical implementations and the later in quantum implementations (since it avoids the exponential cost of post selection).
The performance of this scheme can be improved by further refining the MPO using a variety of cost functions.
b) {\it Improved Classification} can be achieved by a training on the output data, using for example a shallow neural network.
A quantum implementation of a similar scheme can be achieved by introducing ancillae qubits and a stacking unitary $V$ that entangles them with the output.
Training $V$ improves classification.
The procedure is feasible because of the relatively small number of qubits required to label the clusters. An alagous procedure can be applied to the TTO classifier.}
\label{fig:ClassificationAndRefinement}
\end{figure}

\vspace{0.1in}
{\it Quantum Training on Classifier Outputs:}
\noindent 
Similar benefits from training on the classifier output can be achieved in the quantum circuit by a process of data re-uploading\cite{perez2020data,schuld2021effect}. A circuit that achieves this is shown in Fig.~\ref{fig:ClassificationAndRefinement} b). Taking advantage of the classical freedom to copy data, the projection of an image to state on the label subspace is encoded multiple times on the quantum circuit. This may be over additional qubits as illustrated in Fig.~\ref{fig:ClassificationAndRefinement} b) or in circumstances where a subset of qubits can be measured and reset (for example on ion-trap machines), multiple times on the same set of qubits in the course of the computation. 

The essence of this procedure is as follows: 
We define label-space states $\ket{\phi}$ as the projection of an image $\ket{image} $ onto the label space; 
\begin{equation}
\ket{\phi} = \bra{0}^{\otimes(N-L)} U \ket{image}.
\label{eq:label_space_images}
\end{equation}
Label-space states
$\ket{\phi_l}$ and $\ket{\phi_k}$ in different classes are not orthogonal, limiting the ability to distinguish them by quantum means. The states $\ket{\phi_l}^{\otimes M}$ are orthogonal as $M\rightarrow \infty$, thus allowing them to be distinguished in this larger space. Ultimately, every point in the original label Hilbert space becomes orthogonal in the enlarged space; we effectively create a classical state on the label-space, whose position in that space is not subject to the uncertainty of the quantum state.
The procedure of data re-uploading is similar to a multi-class amplitude amplification. Tracing over the ancillae gives the additional non-linearity. As noted in the introduction, this approach is in accord with the interpretation of quantum classification as a kernel method\cite{havlicek_supervised_2019,schuld2021supervised, jerbi2021quantum}

\vspace{0.1in}
\noindent
{\it Constructing }$V$:
As indicated in Fig.~\ref{fig:ClassificationAndRefinement} b), a unitary $V$ is constructed that gives a weighted sum of the predictions on each of the uploaded image data and assigns a class according to 
\begin{equation}
k = \max_l \bra{\phi}^{\otimes M} V^{\dagger} \left[ \mathbb{I}^{\otimes M-1} \otimes \ket{{\bf b}_l}\bra{{\bf b}_l} \right] V \ket{\phi}^{\otimes M}.
 \label{eq:ReuploadClassification}  
\end{equation}
We construct several different approximations to $V$, the first of which attempts a direct analytical approximation to the solution of Eq.(\ref{eq:ReuploadClassification}).
Ideally, the result is a perfect classification of the state:
\begin{equation}\label{eq:PerfectClassification}
   \delta_{l,l_\phi} = \lim_{M \rightarrow \infty} \bra{\phi}^{\otimes M} V^{\dagger} \left[ \mathbb{I}^{\otimes M-1} \otimes \ket{{\bf b}_l}\bra{{\bf b}_l} \right] V \ket{\phi}^{\otimes M},
\end{equation}
where $M - 1$ copies of the label qubits are traced over whilst $\ket{{\bf b}_l}\bra{{\bf b}_l}$ projects $V \ket{\phi}^{\otimes M}$ onto the bitstring representing class $l$.
In the limit of an infinite copies, the states $\ket{\phi}^{\otimes M}$ become orthogonal and Eq.(\ref{eq:PerfectClassification})  can be re-written as
\begin{equation}
   \sum_{\phi} \ket{\phi}\bra{\phi}^{\otimes M} \delta_{l,l_\phi} 
   \approx 
   V^{\dagger} \left[ \mathbb{I}^{\otimes M-1} \otimes \ket{{\bf b}_l}\bra{{\bf b}_l} \right] V.
   \label{eq:InitialiseV}
\end{equation}
This equation can be used to find a good initialisation of $V$ with a finite number of copies of the data.

The procedure is as follows: The tensors $V$ and $W$ are elements of $SU(2^{N_{\text{label}} \times M})$. It is convenient to divide these tensors as a 
$\left(2^{N_{\text{label}}} \; \times \; 2^{N_{\text{label}}}\right)$ grid of $\left(2^{(M-1)N_{\text{label}}}\; \times \; 2^{(M-1)N_{\text{label}}}\right)$ blocks.
The factor of $\left[ \mathbb{I}^{\otimes M-1} \otimes \ket{{\bf b}_l}\bra{{\bf b}_l} \right]$ ensures that for each class $l$ only the $l^{th}$ column of this grid contributes. Taking a singular value decomposition, $W_l \Lambda_l W_l^{\dagger}$, of the left-hand side of Eq.(\ref{eq:InitialiseV}) for the $l^{th}$ image class, the $l^{th}$ column of $V$ can be determined by identifying $V_l=\sqrt{\Lambda_l} W^\dagger$,where only the largest $2^{(M-1)N_{\text{label}}}$
singular values of $\Lambda_l$ are retained. 
Finally, all columns of V are made to be orthogonal {\it via} a polar decomposition. 

\noindent
{\it Results}:
The quantum stacking/refinement procedure works well. As shown in Table \ref{tab:RefinementResults}, the procedure above improves performance even when only a single copy of the data is uploaded. With $3$ copies of the data, the results of this deterministic initialisation out-perform classical training on the output of the quantum circuit with 170 parameters trained over 10,000 epochs.

\begin{table*}[t]
\begin{center}
\begin{tabular}{|c|cc|cc|}
\hline
\multirow{2}{*}{} & \multicolumn{2}{c|}{MNIST}                                                                                                                           & \multicolumn{2}{c|}{Fashion MNIST}                                                                                                                   \\ \cline{2-5} 
                  & \multicolumn{1}{c|}{\begin{tabular}[c]{@{}c@{}}Training Accuracy\\ (\%)\end{tabular}} & \begin{tabular}[c]{@{}c@{}}Test Accuracy\\ (\%)\end{tabular} & \multicolumn{1}{c|}{\begin{tabular}[c]{@{}c@{}}Training Accuracy\\ (\%)\end{tabular}} & \begin{tabular}[c]{@{}c@{}}Test Accuracy\\ (\%)\end{tabular} \\ \hline
Initial           & \multicolumn{1}{c|}{76.44}                                                            & 78.85                                                        & \multicolumn{1}{c|}{70.50}                                                            & 69.75                                                        \\ \hline
Classical         & \multicolumn{1}{c|}{82.27}                                                            & 83.92                                                        & \multicolumn{1}{c|}{79.22}                                                            & 77.52                                                        \\ \hline
1 Copy            & \multicolumn{1}{c|}{76.71}                                                            & 79.13                                                        & \multicolumn{1}{c|}{71.59}                                                            & 70.57                                                        \\ \hline
2 Copies          & \multicolumn{1}{c|}{81.61}                                                            & 83.63                                                        & \multicolumn{1}{c|}{74.60}                                                            & 73.30                                                        \\ \hline
3 Copies          & \multicolumn{1}{c|}{83.19}                                                            & 84.86                                                        & \multicolumn{1}{c|}{75.67}                                                            & 74.59                                                        \\ \hline
\end{tabular}
\end{center}
\caption{{\bf Test accuracies from stacking refinement of} $D_{\text{total}} = 32$ {\bf MPO classifiers}.
The results of training on the outputs  --- known classically as stacking ---
are shown.
The classical procedure trains a single dense layer neural network on the outputs of  Fig.~\ref{fig:ClassificationAndRefinement}a) using a cross entropy cost function over 10,000 epochs.
Quantum stacking uses a data re-uploading scheme, dependent upon the number of copies of the label qubits uploaded to the quantum circuit. 
It is notable that the quantum procedure of data re-uploading {\it and its deterministic initialisation} out performs the classical procedure after only uploading three copies for MNIST.}
\label{tab:RefinementResults}
\end{table*}

\begin{table*}[t]

\vspace{0.5cm}
\begin{center}
\begin{tabular}{|c|cc|cc|}
\hline
\multirow{2}{*}{} & \multicolumn{2}{c|}{MNIST - TTN}                                                                                                                           & \multicolumn{2}{c|}{Fashion MNIST - TTN}                                                                                                                   \\ \cline{2-5} 
                  & \multicolumn{1}{c|}{\begin{tabular}[c]{@{}c@{}}Training Accuracy\\ (\%)\end{tabular}} & \begin{tabular}[c]{@{}c@{}}Test Accuracy\\ (\%)\end{tabular} & \multicolumn{1}{c|}{\begin{tabular}[c]{@{}c@{}}Training Accuracy\\ (\%)\end{tabular}} & \begin{tabular}[c]{@{}c@{}}Test Accuracy\\ (\%)\end{tabular} \\ \hline
Initial           & \multicolumn{1}{c|}{78.17}                                                            & 80.09                                                       & \multicolumn{1}{c|}{71.62}                                                            & 70.1                                                            \\ \hline
2 Copies            & \multicolumn{1}{c|}{83.47}                                                            & 85.07                                                        & \multicolumn{1}{c|}{72.60}                                                            & 71.55                                                        \\ \hline
3 Copies          & \multicolumn{1}{c|}{84.63}                                                            & 86.16                                                        & \multicolumn{1}{c|}{73.17}                                                            & 72.33                                                      \\ \hline

\end{tabular}
\end{center}
\caption{{\bf Test accuracies from refinement of} $D_{\text{total}} = 16$ {\bf TTO classifiers}.
Shows the performance of the quantum stacking/data re-uploading scheme depending upon the number of copies of the image data uploaded to the quantum circuit. Results are similar to Table~\ref{tab:RefinementResults} but the TTN appears to slightly outperform the MPS decomposition on MNIST but performs worse on fashion MNIST.}
\label{tab:RefinementResultsTree}
\end{table*}

\subsubsection{Improving Quantum Stacking Efficiency} \label{ssec:efficiency}

The primary bottleneck in the above method is the construction of the stacking unitary, $V$, in Eq.~\ref{eq:InitialiseV}. The number of parameters of $V$ grows exponentially with $M$, the number of times that the image data is uploaded. To remedy this we propose two different factorisations that allow stacking to be carried out in a scalable manner.\\

\begin{figure}[ht]
    \includegraphics[width = 0.45\textwidth]{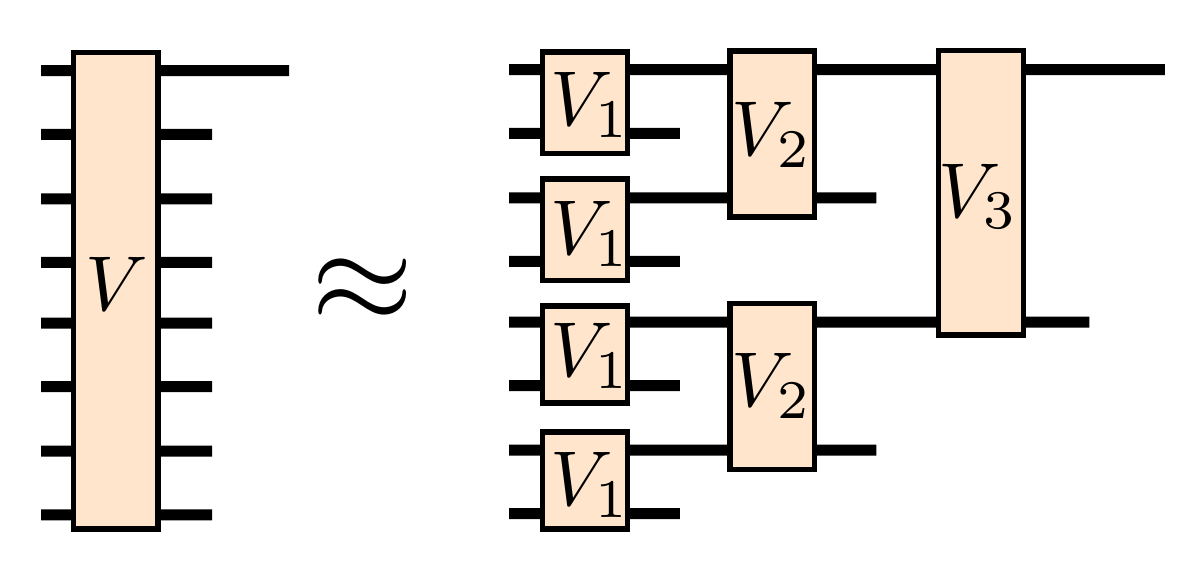}
    \caption{{\textbf{Hierarchical Stacking Circuit:}} Hierarchical stacking is achieved by factorising the stacking unitary $V$ into layers. We illustrate a stacking that uploads eight copies of the data. $V$ combines eight copies of the label space onto one copy. It is factorised combining two copies in each layer over a total of three layers. The resulting circuit naturally has a tree structure. The unitaries in each layer can be found by iterative solution of Eq.(\ref{eq:InitialiseV}) using the input from each layer on the left hand side. }
    \label{fig:hierarchical_stacking_circuit}
\end{figure}

{\it Hierarchical Stacking}:
The first method is to construct $V$ hierarchically. We construct a layered factorisation as depicted in Fig.~\ref{fig:hierarchical_stacking_circuit}, where in each each layer a stacking unitary $V_n$ combines a small number of copies of the label space. The unitaries at each layer are obtained by solving Eq.~\ref{eq:InitialiseV}, using the outputs from the previous layer to construct the left hand side. In this way the number of copies (and the number of qubits) rises exponentially with the number of layers, but the individual unitaries are of a manageable size. 

\begin{figure}[h]
\centering
\includegraphics[width=\columnwidth]{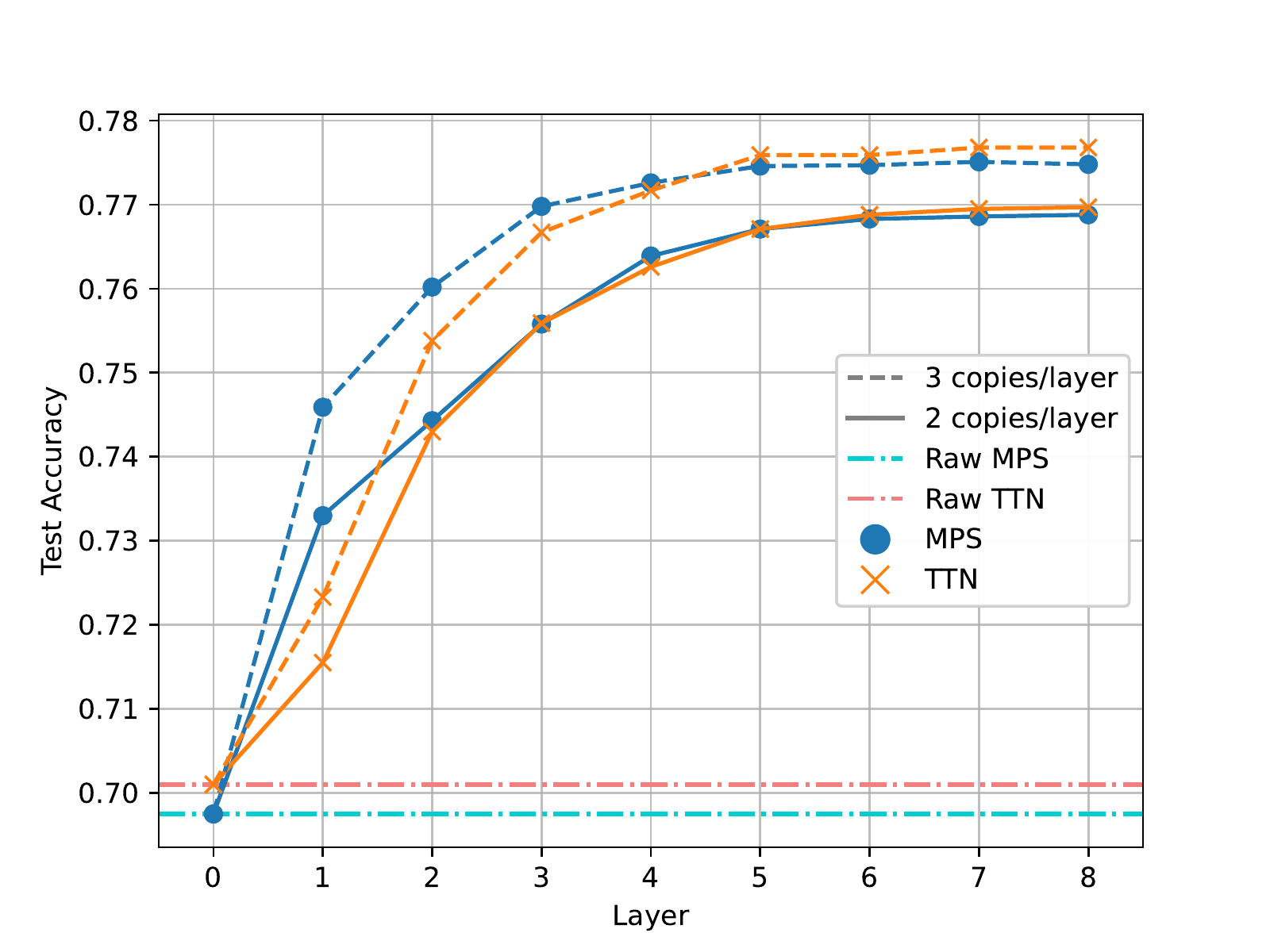}
\caption{{\bf Hierarchical Quantum Stacking for fashion MNIST.}
Hierarchical stacking resulting from sequential solutions of Eq.~\ref{eq:InitialiseV} for a small number of copies in each $V_i$. Layer $0$ corresponds to the raw accuracy for the original MPO/TTO classifier for fashion MNIST. As we increase the number of layers we see a monotonic accuracy improvement for both the MPS and TTN decompositions. Furthermore, increasing the number of copies (from $2$ to $3$) into each $V_i$ also improves accuracy.
}
\label{fig:hierarchical_stacking_fashion}
\end{figure}

The results of this procedure are shown in Fig.~\ref{fig:hierarchical_stacking_fashion} for the fashion MNIST dataset. Here, we achieve almost  $10\%$ improvement in classification accuracy over the raw MPO/TTO classification. The test accuracy saturates after $4$ or $5$ layers and we note that increasing the number copies inputted to each unitary, $V_n$, from $2$ to $3$ increases the overall accuracy by about $1\%$. We observe similar features in applying this for MNIST, whose results can be seen in Appendix~\ref{app:quantum_stacking}. 

\vspace{-1cm}
{\it Tensor Network Stacking}:
Anticipating that increasing the number of copies in a single layer stacking might lead to a more rapid increase in test accuracy motivates our second factorisation of $V$ - essentially as an MPO.
Constructing $V$  using our direct approximation to the solution of Eq.(\ref{eq:ReuploadClassification}) becomes prohibitively costly beyond three copies of the data. It remains to be seen whether there is an efficient way to construct an MPO approximation to the solution of this equation.
Instead
we repeat the construction of the MPO classifier $U$ but on multiple copies of the label states rather than the original amplitude encoded image. In Fig.~\ref{fig:MPSEncoding}, each raw image is amplitude encoded as a tensor network state. In order to modify this procedure for stacking, we replace the amplitude encoded images with a tensor product of the label states, $\ket{\psi_l}^{\otimes M}$, and repeat the tensor network decomposition and batch adding to form a second MPO classifier. In this way, by regulating the bond dimensions of the encoded images and tensor network operator classifier, we can build a stacking unitary with only a linear cost in $M$. 

\begin{figure}[h!]
\includegraphics[width=0.85\columnwidth]{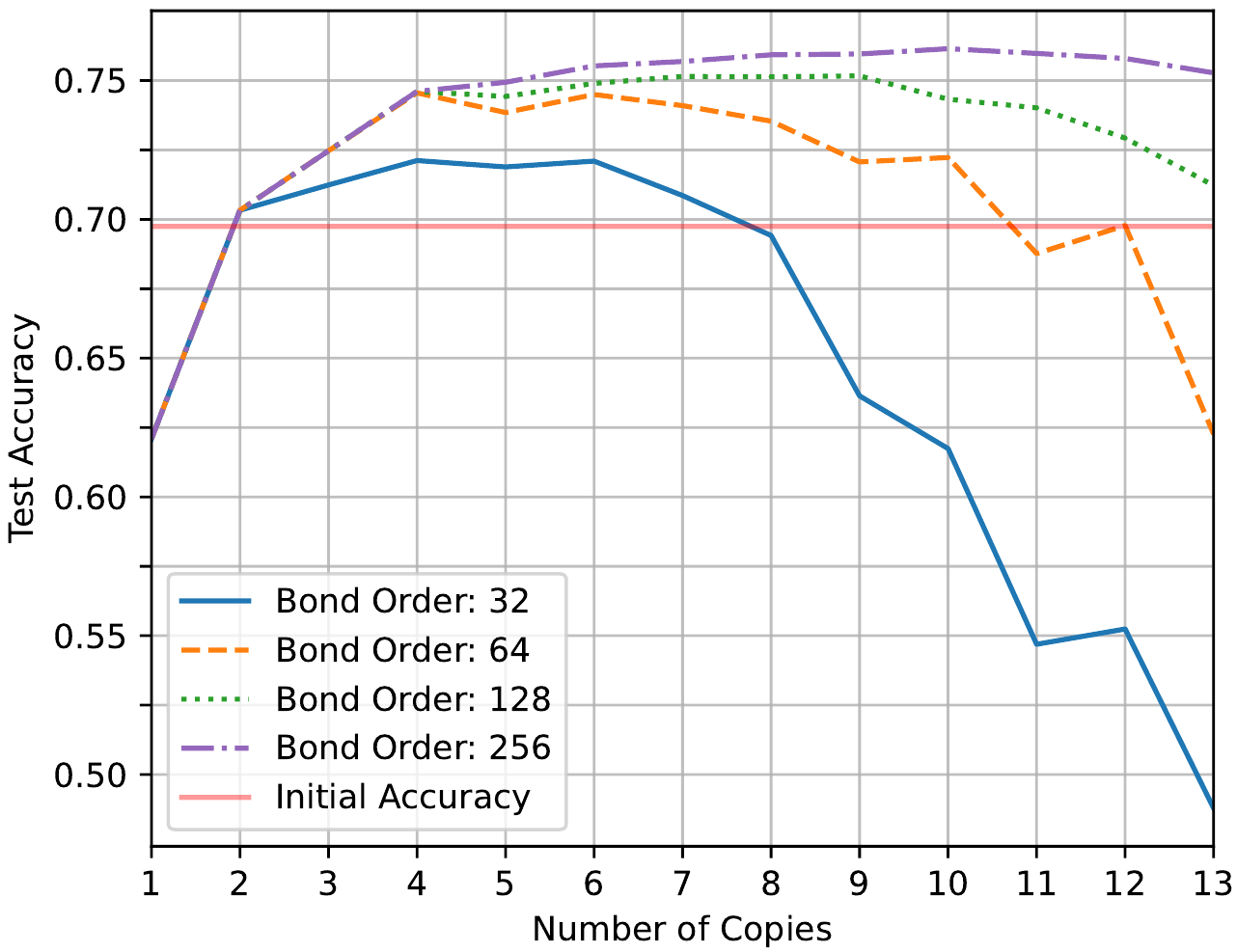}
\caption{
{\bf Tensor Network Quantum Stacking for fashion MNIST.} The test accuracy of quantum stacking as a function of the number of copies of the data uploaded and the bond order of the tensor network approximations to the stacking unitary $V$. Increasing the number of copies leads to an increase in performance- regulated by bond order. $V$ is constructed from the batch addition of the copied training image predictions encoded as MPS as described in the main text. 
.}
\label{fig:mps_stacking}
\end{figure}

The results of this procedure are shown in Fig.~\ref{fig:mps_stacking}. It is evident that increasing the number of copies combined in the single layer stacking procedure can yield significant improvement in classification, provided that the bond order of the MPO is not truncated too far. We note that when using just one or two copies, an approximation to $V$ produced by the singular-value decomposition procedure actually performs worse than doing no stacking at all. This is in contrast to the result obtained using Eq.~\ref{eq:InitialiseV} which is much closer to a direct analytical optimisation of a cost function related to the classification. 

Which approach is most favourable is likely both device and data dependent. The tensor network representation of the stacking unitary - and its construction by sequential singular value decompositions - can also be carried out in a hierarchical way. One might speculate upon the possibility of a \emph{direct} MPS solution of Eq.~\ref{eq:InitialiseV}. An exploration of this is beyond the scope of this work. Our results serve to demonstrate that stacking, facilitated by multiple uploadings of classical data can significantly enhance the performance of tensor network-based quantum classification. 

\section{Discussion}
\label{sec:Discussion}

We have demonstrated a deterministic initialisation and refinement of MPO/TTOs to classify MPS/TTN representations of amplitude encoded image data. The two steps in our scheme --- the initialisation of $U$ and then $V$ --- may be considered a feature selection encoded on the label space followed by a classification based on those features\cite{schuld2019machine,schuld2019quantum,lloyd2020quantum, larose_robust_2020}. Our quantum version of the refinement scheme uses data re-uploading to achieve the same results as classical stacking or training on the output data.

Alternative, deterministic schemes developed in Refs.\cite{stoudenmire2016supervised,huggins2019towards,reyes2021multi,stoudenmire2018learning} use product state image encodings, together with MPS or TTN classifiers. The latter are optimised by a DMRG-like scheme\cite{SteveWhiteDMRG} with a particular choice of cost function.
These classifiers are not orthogonalised for implementation on a quantum circuit and it is not clear whether doing so would degrade performance. Moreover, there are interesting differences in performance and trade-offs in encoding and classification between this and our approach that bear further investigation. In our work we have focused upon faithful encoding of pixel intensities to the minimum number of qubits. There may be advantages in encodings that use fewer qubits -  For example, an intermediate between product state and MPS encoding could preserve some locality within the data whilst still efficiently capturing entanglement on fewer qubits\cite{schuld2021effect,dilip_data_2022}. Our quantum refinement/stacking procedure uses a similar relaxation of the number of qubits, albeit increasing the number of qubits after the feature extraction rather than in the encoding. It would be interesting to determine when a relaxation of the encoding outperforms stacking using  data re-uploading and {\it vice versa}, and whether there is an advantage in the combination the two. 

It can readily be shown that linear models, such as a single layer linear perceptron, have closed-form solutions and therefore can also be classified as deterministic. However, such models do not perform as well as the scheme presented here. Additionally, analytic solutions to the linear models require inverting a matrix with size that scales linearly with the number of training images and features. This is computationally expensive for large data sets when compared to our scheme which does not suffer from this scaling.

Training of quantum classifiers can be challenging due the existence of barren plateaux\cite{mcclean2018barren,holmes2021connecting,marrero2021entanglement,patti2021entanglement,wang2021noise} --- regions of parameter space where cost functions are exponentially flat (or polynomially flat in convolutional networks). Good initialisations are essential. The structure of MPS encoded data allows deterministic initialisations that perform very well\cite{dborin2021matrix}.

Quantum advantage has been identified in the simulation of quantum systems\cite{Lin2021real} and in the classification of quantum data\cite{huang2021quantum}. Similar advantage in the classification of classical data is harder to identify\cite{huang2021power, huang2021information, cotler2021revisiting} --- its existence depends upon the structure of the data to be classified. Mapping the correlations of classical data to the entanglement structure of an encoded state brings the possibility (with the appropriate mapping) of a quantum advantage in its manipulation. Due to the loss of locality in the amplitude encoding, the physical representation of the encoded entanglement structure is less obvious. We suspect the structure to be more global, similar to that found in principle component analysis where less important correlations within the image are given a reduced weighting.

We have chosen a label space that permits one orthogonal bitstring for each image class. However, a data uploading scheme can potentially make a classification when image data is compressed to a smaller space\cite{perez2020data}. This corresponds to extracting fewer features in the data than image classes and relying upon different values for these features to separate states. Such a scheme requires fewer parameters though potentially more copies of the image to be uploaded to get good identification. Ultimately, this reduction in the size of the label space is limited by fidelity of the device.

Tensor networks create a bridge between classical and quantum machine learning.
By using this technology we can benefit from the advantages of both. We have shown how the algebraic structure of tensor network states and the geometry of the Hilbert space allow excellent initialisation of unitaries for feature extraction and systematic ways to use data re-uploading in a stacking procedure to refine the resultant classification. This requires only the tools of linear algebra - the singular value decompositions. It is possible that quantum advantage in performing such linear algebraic tasks\cite{harrow_quantum_2009, dervovic_quantum_2018} could yet lead to advantage in the training procedure itself and not only the application of the machine learning model. 

 \section{Acknowledgements}  We would like to thank Ehud Altman, Yimu Bao and Soonwon Choi for early discussions on related questions and acknowledge a suggestion from Frank Pollmann of the potential utility of an MPS representation of amplitude encodings.

Note: At a late stage of manuscript preparation we became aware of Ref.\cite{dilip_data_2022}, which contains some similar ideas regarding the translation and compression of image data and tensor network classifiers onto quantum computers for quantum machine learning.

\bibliographystyle{naturemag}
\bibliography{Bibliography.bib}

\begin{thebibliography}{10}
\expandafter\ifx\csname url\endcsname\relax
  \def\url#1{\texttt{#1}}\fi
\expandafter\ifx\csname urlprefix\endcsname\relax\def\urlprefix{URL }\fi
\providecommand{\bibinfo}[2]{#2}
\providecommand{\eprint}[2][]{\url{#2}}

\bibitem{nielsen_chuang}
\bibinfo{author}{Nielsen, M.~A.} \& \bibinfo{author}{Chuang, I.~L.}
\newblock \emph{\bibinfo{title}{Quantum Computation and Quantum Information:
  10th Anniversary Edition}} (\bibinfo{publisher}{Cambridge University Press},
  \bibinfo{address}{USA}, \bibinfo{year}{2011}), \bibinfo{edition}{10th} edn.
\newblock
  \urlprefix\url{https://www.cambridge.org/highereducation/books/quantum-computation-and-quantum-information/01E10196D0A682A6AEFFEA52D53BE9AE#overview}.

\bibitem{farhi2018classification}
\bibinfo{author}{Farhi, E.} \& \bibinfo{author}{Neven, H.}
\newblock \bibinfo{title}{Classification with {Quantum} {Neural} {Networks} on
  {Near} {Term} {Processors}}.
\newblock \emph{\bibinfo{journal}{arXiv:1802.06002 [quant-ph]}}
  (\bibinfo{year}{2018}).
\newblock \urlprefix\url{http://arxiv.org/abs/1802.06002}.
\newblock \bibinfo{note}{ArXiv: 1802.06002}.

\bibitem{schuld2019quantum}
\bibinfo{author}{Schuld, M.} \& \bibinfo{author}{Killoran, N.}
\newblock \bibinfo{title}{Quantum machine learning in feature hilbert spaces}.
\newblock \emph{\bibinfo{journal}{Phys. Rev. Lett.}}
  \textbf{\bibinfo{volume}{122}}, \bibinfo{pages}{040504}
  (\bibinfo{year}{2019}).
\newblock
  \urlprefix\url{https://link.aps.org/doi/10.1103/PhysRevLett.122.040504}.

\bibitem{mcclean2018barren}
\bibinfo{author}{McClean, J.~R.}, \bibinfo{author}{Boixo, S.},
  \bibinfo{author}{Smelyanskiy, V.~N.}, \bibinfo{author}{Babbush, R.} \&
  \bibinfo{author}{Neven, H.}
\newblock \bibinfo{title}{Barren plateaus in quantum neural network training
  landscapes}.
\newblock \emph{\bibinfo{journal}{Nature Communications}}
  \textbf{\bibinfo{volume}{9}}, \bibinfo{pages}{4812} (\bibinfo{year}{2018}).
\newblock \urlprefix\url{https://www.nature.com/articles/s41467-018-07090-4}.
\newblock \bibinfo{note}{Number: 1 Publisher: Nature Publishing Group}.

\bibitem{holmes2021connecting}
\bibinfo{author}{Holmes, Z.}, \bibinfo{author}{Sharma, K.},
  \bibinfo{author}{Cerezo, M.} \& \bibinfo{author}{Coles, P.~J.}
\newblock \bibinfo{title}{Connecting ansatz expressibility to gradient
  magnitudes and barren plateaus}.
\newblock \emph{\bibinfo{journal}{PRX Quantum}} \textbf{\bibinfo{volume}{3}},
  \bibinfo{pages}{010313} (\bibinfo{year}{2022}).
\newblock \urlprefix\url{https://link.aps.org/doi/10.1103/PRXQuantum.3.010313}.

\bibitem{marrero2021entanglement}
\bibinfo{author}{Ortiz~Marrero, C.}, \bibinfo{author}{Kieferov\'a, M.} \&
  \bibinfo{author}{Wiebe, N.}
\newblock \bibinfo{title}{Entanglement-induced barren plateaus}.
\newblock \emph{\bibinfo{journal}{PRX Quantum}} \textbf{\bibinfo{volume}{2}},
  \bibinfo{pages}{040316} (\bibinfo{year}{2021}).
\newblock \urlprefix\url{https://link.aps.org/doi/10.1103/PRXQuantum.2.040316}.

\bibitem{patti2021entanglement}
\bibinfo{author}{Patti, T.~L.}, \bibinfo{author}{Najafi, K.},
  \bibinfo{author}{Gao, X.} \& \bibinfo{author}{Yelin, S.~F.}
\newblock \bibinfo{title}{Entanglement devised barren plateau mitigation}.
\newblock \emph{\bibinfo{journal}{Phys. Rev. Research}}
  \textbf{\bibinfo{volume}{3}}, \bibinfo{pages}{033090} (\bibinfo{year}{2021}).
\newblock
  \urlprefix\url{https://link.aps.org/doi/10.1103/PhysRevResearch.3.033090}.

\bibitem{wang2021noise}
\bibinfo{author}{Wang, S.} \emph{et~al.}
\newblock \bibinfo{title}{Noise-induced barren plateaus in variational quantum
  algorithms}.
\newblock \emph{\bibinfo{journal}{Nature Communications}}
  \textbf{\bibinfo{volume}{12}}, \bibinfo{pages}{6961} (\bibinfo{year}{2021}).
\newblock \urlprefix\url{https://www.nature.com/articles/s41467-021-27045-6}.

\bibitem{dborin2021matrix}
\bibinfo{author}{Dborin, J.}, \bibinfo{author}{Barratt, F.},
  \bibinfo{author}{Wimalaweera, V.}, \bibinfo{author}{Wright, L.} \&
  \bibinfo{author}{Green, A.~G.}
\newblock \bibinfo{title}{Matrix {Product} {State} {Pre}-{Training} for
  {Quantum} {Machine} {Learning}}.
\newblock \emph{\bibinfo{journal}{arXiv:2106.05742 [quant-ph]}}
  (\bibinfo{year}{2021}).
\newblock \urlprefix\url{http://arxiv.org/abs/2106.05742}.

\bibitem{preskill2018quantum}
\bibinfo{author}{Preskill, J.}
\newblock \bibinfo{title}{Quantum {Computing} in the {NISQ} era and beyond}.
\newblock \emph{\bibinfo{journal}{Quantum}} \textbf{\bibinfo{volume}{2}},
  \bibinfo{pages}{79} (\bibinfo{year}{2018}).
\newblock \urlprefix\url{https://quantum-journal.org/papers/q-2018-08-06-79/}.
\newblock \bibinfo{note}{Publisher: Verein zur F{\"o}rderung des Open Access
  Publizierens in den Quantenwissenschaften}.

\bibitem{schollwock2011density}
\bibinfo{author}{Schollw{\"{o}}ck, U.}
\newblock \bibinfo{title}{The density-matrix renormalization group in the age
  of matrix product states}.
\newblock \emph{\bibinfo{journal}{Ann. Phys. (NY)}}
  \textbf{\bibinfo{volume}{326}}, \bibinfo{pages}{96 -- 192}
  (\bibinfo{year}{2011}).
\newblock
  \urlprefix\url{http://www.sciencedirect.com/science/article/pii/S0003491610001752}.
\newblock \bibinfo{note}{January 2011 Special Issue}.

\bibitem{orus_review}
\bibinfo{author}{Or{\'{u}}s, R.}
\newblock \bibinfo{title}{A practical introduction to tensor networks: Matrix
  product states and projected entangled pair states}.
\newblock \emph{\bibinfo{journal}{Ann. Phys. (NY)}}
  \textbf{\bibinfo{volume}{349}}, \bibinfo{pages}{117 -- 158}
  (\bibinfo{year}{2014}).
\newblock
  \urlprefix\url{http://www.sciencedirect.com/science/article/pii/S0003491614001596}.

\bibitem{Perez-Garcia2006}
\bibinfo{author}{Perez-Garcia, D.}, \bibinfo{author}{Verstraete, F.},
  \bibinfo{author}{Wolf, M.~M.} \& \bibinfo{author}{Cirac, J.~I.}
\newblock \bibinfo{title}{{Matrix product state representations}}.
\newblock \emph{\bibinfo{journal}{arXiv:quant-ph/0608197v2}}
  \bibinfo{pages}{1--28} (\bibinfo{year}{2006}).
\newblock \urlprefix\url{http://arxiv.org/abs/quant-ph/0608197}.
\newblock \eprint{0608197v2}.

\bibitem{shi_classical_2006}
\bibinfo{author}{Shi, Y.-Y.}, \bibinfo{author}{Duan, L.-M.} \&
  \bibinfo{author}{Vidal, G.}
\newblock \bibinfo{title}{Classical simulation of quantum many-body systems
  with a tree tensor network}.
\newblock \emph{\bibinfo{journal}{Phys. Rev. A}} \textbf{\bibinfo{volume}{74}},
  \bibinfo{pages}{022320} (\bibinfo{year}{2006}).
\newblock \urlprefix\url{https://link.aps.org/doi/10.1103/PhysRevA.74.022320}.

\bibitem{stoudenmire2016supervised}
\bibinfo{author}{Stoudenmire, E.} \& \bibinfo{author}{Schwab, D.~J.}
\newblock \bibinfo{title}{Supervised {Learning} with {Tensor} {Networks}}.
\newblock In \bibinfo{editor}{Lee, D.}, \bibinfo{editor}{Sugiyama, M.},
  \bibinfo{editor}{Luxburg, U.}, \bibinfo{editor}{Guyon, I.} \&
  \bibinfo{editor}{Garnett, R.} (eds.) \emph{\bibinfo{booktitle}{Advances in
  {Neural} {Information} {Processing} {Systems}}}, vol.~\bibinfo{volume}{29}
  (\bibinfo{publisher}{Curran Associates, Inc.}, \bibinfo{year}{2016}).
\newblock
  \urlprefix\url{https://proceedings.neurips.cc/paper/2016/file/5314b9674c86e3f9d1ba25ef9bb32895-Paper.pdf}.

\bibitem{huggins2019towards}
\bibinfo{author}{Huggins, W.}, \bibinfo{author}{Patil, P.},
  \bibinfo{author}{Mitchell, B.}, \bibinfo{author}{Whaley, K.~B.} \&
  \bibinfo{author}{Stoudenmire, E.~M.}
\newblock \bibinfo{title}{Towards quantum machine learning with tensor
  networks}.
\newblock \emph{\bibinfo{journal}{Quantum Science and Technology}}
  \textbf{\bibinfo{volume}{4}}, \bibinfo{pages}{024001} (\bibinfo{year}{2019}).
\newblock \urlprefix\url{https://doi.org/10.1088/2058-9565/aaea94}.
\newblock \bibinfo{note}{Publisher: IOP Publishing}.

\bibitem{novikov2016exponential}
\bibinfo{author}{Novikov, A.}, \bibinfo{author}{Trofimov, M.} \&
  \bibinfo{author}{Oseledets, I.}
\newblock \bibinfo{title}{Exponential machines}.
\newblock \emph{\bibinfo{journal}{arXiv preprint arXiv:1605.03795}}
  (\bibinfo{year}{2016}).

\bibitem{schon2005sequential}
\bibinfo{author}{Sch{\"o}n, C.}, \bibinfo{author}{Solano, E.},
  \bibinfo{author}{Verstraete, F.}, \bibinfo{author}{Cirac, J.~I.} \&
  \bibinfo{author}{Wolf, M.~M.}
\newblock \bibinfo{title}{Sequential generation of entangled multiqubit
  states}.
\newblock \emph{\bibinfo{journal}{Physical review letters}}
  \textbf{\bibinfo{volume}{95}}, \bibinfo{pages}{110503}
  (\bibinfo{year}{2005}).
\newblock \urlprefix\url{https://doi.org/10.1103/PhysRevLett.95.110503}.

\bibitem{banuls2008sequentially}
\bibinfo{author}{Banuls, M.-C.}, \bibinfo{author}{P{\'e}rez-Garc{\'\i}a, D.},
  \bibinfo{author}{Wolf, M.~M.}, \bibinfo{author}{Verstraete, F.} \&
  \bibinfo{author}{Cirac, J.~I.}
\newblock \bibinfo{title}{Sequentially generated states for the study of
  two-dimensional systems}.
\newblock \emph{\bibinfo{journal}{Physical Review A}}
  \textbf{\bibinfo{volume}{77}}, \bibinfo{pages}{052306}
  (\bibinfo{year}{2008}).

\bibitem{wei2022sequential}
\bibinfo{author}{Wei, Z.-Y.}, \bibinfo{author}{Malz, D.} \&
  \bibinfo{author}{Cirac, J.~I.}
\newblock \bibinfo{title}{Sequential generation of projected entangled-pair
  states}.
\newblock \emph{\bibinfo{journal}{Physical Review Letters}}
  \textbf{\bibinfo{volume}{128}}, \bibinfo{pages}{010607}
  (\bibinfo{year}{2022}).

\bibitem{hierarchical_quantum_classifiers}
\bibinfo{author}{Grant, E.} \emph{et~al.}
\newblock \bibinfo{title}{Hierarchical quantum classifiers}.
\newblock \emph{\bibinfo{journal}{npj Quantum Inf.}}
  \textbf{\bibinfo{volume}{4}}, \bibinfo{pages}{65} (\bibinfo{year}{2018}).
\newblock \urlprefix\url{https://doi.org/10.1038/s41534-018-0116-9}.

\bibitem{huang2021quantum}
\bibinfo{author}{Huang, H.-Y.} \emph{et~al.}
\newblock \bibinfo{title}{Quantum advantage in learning from experiments}.
\newblock \emph{\bibinfo{journal}{arXiv:2112.00778}}  (\bibinfo{year}{2021}).
\newblock \urlprefix\url{http://arxiv.org/abs/2112.00778}.

\bibitem{wolpert1992stacked}
\bibinfo{author}{Wolpert, D.~H.}
\newblock \bibinfo{title}{Stacked generalization}.
\newblock \emph{\bibinfo{journal}{Neural networks}}
  \textbf{\bibinfo{volume}{5}}, \bibinfo{pages}{241--259}
  (\bibinfo{year}{1992}).

\bibitem{perez2020data}
\bibinfo{author}{P{\'e}rez-Salinas, A.}, \bibinfo{author}{Cervera-Lierta, A.},
  \bibinfo{author}{Gil-Fuster, E.} \& \bibinfo{author}{Latorre, J.~I.}
\newblock \bibinfo{title}{Data re-uploading for a universal quantum
  classifier}.
\newblock \emph{\bibinfo{journal}{Quantum}} \textbf{\bibinfo{volume}{4}},
  \bibinfo{pages}{226} (\bibinfo{year}{2020}).
\newblock \urlprefix\url{https://quantum-journal.org/papers/q-2020-02-06-226/}.
\newblock \bibinfo{note}{Publisher: Verein zur F{\"o}rderung des Open Access
  Publizierens in den Quantenwissenschaften}.

\bibitem{schuld2021effect}
\bibinfo{author}{Schuld, M.}, \bibinfo{author}{Sweke, R.} \&
  \bibinfo{author}{Meyer, J.~J.}
\newblock \bibinfo{title}{Effect of data encoding on the expressive power of
  variational quantum-machine-learning models}.
\newblock \emph{\bibinfo{journal}{Phys. Rev. A}}
  \textbf{\bibinfo{volume}{103}}, \bibinfo{pages}{032430}
  (\bibinfo{year}{2021}).
\newblock \urlprefix\url{https://link.aps.org/doi/10.1103/PhysRevA.103.032430}.

\bibitem{caro2021encoding}
\bibinfo{author}{Caro, M.~C.}, \bibinfo{author}{Gil-Fuster, E.},
  \bibinfo{author}{Meyer, J.~J.}, \bibinfo{author}{Eisert, J.} \&
  \bibinfo{author}{Sweke, R.}
\newblock \bibinfo{title}{Encoding-dependent generalization bounds for
  parametrized quantum circuits}.
\newblock \emph{\bibinfo{journal}{Quantum}} \textbf{\bibinfo{volume}{5}},
  \bibinfo{pages}{582} (\bibinfo{year}{2021}).
\newblock \urlprefix\url{https://quantum-journal.org/papers/q-2021-11-17-582/}.
\newblock \bibinfo{note}{Publisher: Verein zur F{\"o}rderung des Open Access
  Publizierens in den Quantenwissenschaften}.

\bibitem{long2021properties}
\bibinfo{author}{Long, P.~M.}
\newblock \bibinfo{title}{Properties of the after kernel}.
\newblock \emph{\bibinfo{journal}{arXiv preprint arXiv:2105.10585}}
  (\bibinfo{year}{2021}).

\bibitem{havlicek_supervised_2019}
\bibinfo{author}{Havl{\'\i}{\v c}ek, V.} \emph{et~al.}
\newblock \bibinfo{title}{Supervised learning with quantum-enhanced feature
  spaces}.
\newblock \emph{\bibinfo{journal}{Nature}} \textbf{\bibinfo{volume}{567}},
  \bibinfo{pages}{209--212} (\bibinfo{year}{2019}).
\newblock \urlprefix\url{https://www.nature.com/articles/s41586-019-0980-2}.
\newblock \bibinfo{note}{Number: 7747 Publisher: Nature Publishing Group}.

\bibitem{schuld2021supervised}
\bibinfo{author}{Schuld, M.}
\newblock \bibinfo{title}{Supervised quantum machine learning models are kernel
  methods}.
\newblock \emph{\bibinfo{journal}{arXiv:2101.11020}}  (\bibinfo{year}{2021}).
\newblock \urlprefix\url{http://arxiv.org/abs/2101.11020}.
\newblock \bibinfo{note}{ArXiv:2101.11020 [quant-ph, stat] type: article}.

\bibitem{jerbi2021quantum}
\bibinfo{author}{Jerbi, S.} \emph{et~al.}
\newblock \bibinfo{title}{Quantum machine learning beyond kernel methods}.
\newblock \emph{\bibinfo{journal}{arXiv:2110.13162 [quant-ph, stat]}}
  (\bibinfo{year}{2022}).
\newblock \urlprefix\url{http://arxiv.org/abs/2110.13162}.
\newblock \bibinfo{note}{ArXiv: 2110.13162}.

\bibitem{haug_large-scale_2021}
\bibinfo{author}{Haug, T.}, \bibinfo{author}{Self, C.~N.} \&
  \bibinfo{author}{Kim, M.~S.}
\newblock \bibinfo{title}{Large-scale quantum machine learning}.
\newblock \emph{\bibinfo{journal}{arXiv:2108.01039 [quant-ph, stat]}}
  (\bibinfo{year}{2021}).
\newblock \urlprefix\url{http://arxiv.org/abs/2108.01039}.
\newblock \bibinfo{note}{ArXiv: 2108.01039}.

\bibitem{li_quantum_2021}
\bibinfo{author}{Li, W.}, \bibinfo{author}{Lu, S.} \& \bibinfo{author}{Deng,
  D.-L.}
\newblock \bibinfo{title}{Quantum federated learning through blind quantum
  computing}.
\newblock \emph{\bibinfo{journal}{Science China Physics, Mechanics \&
  Astronomy}} \textbf{\bibinfo{volume}{64}}, \bibinfo{pages}{100312}
  (\bibinfo{year}{2021}).
\newblock \urlprefix\url{http://arxiv.org/abs/2103.08403}.
\newblock \bibinfo{note}{ArXiv: 2103.08403}.

\bibitem{johri_nearest_2021}
\bibinfo{author}{Johri, S.} \emph{et~al.}
\newblock \bibinfo{title}{Nearest centroid classification on a trapped ion
  quantum computer}.
\newblock \emph{\bibinfo{journal}{npj Quantum Information}}
  \textbf{\bibinfo{volume}{7}}, \bibinfo{pages}{1--11} (\bibinfo{year}{2021}).
\newblock \urlprefix\url{https://www.nature.com/articles/s41534-021-00456-5}.
\newblock \bibinfo{note}{Number: 1 Publisher: Nature Publishing Group}.

\bibitem{peters_machine_2021}
\bibinfo{author}{Peters, E.} \emph{et~al.}
\newblock \bibinfo{title}{Machine learning of high dimensional data on a noisy
  quantum processor}.
\newblock \emph{\bibinfo{journal}{npj Quantum Information}}
  \textbf{\bibinfo{volume}{7}}, \bibinfo{pages}{1--5} (\bibinfo{year}{2021}).
\newblock \urlprefix\url{https://www.nature.com/articles/s41534-021-00498-9}.
\newblock \bibinfo{note}{Number: 1 Publisher: Nature Publishing Group}.

\bibitem{marshall_high_2022}
\bibinfo{author}{Marshall, S.~C.}, \bibinfo{author}{Gyurik, C.} \&
  \bibinfo{author}{Dunjko, V.}
\newblock \bibinfo{title}{High {Dimensional} {Quantum} {Machine} {Learning}
  {With} {Small} {Quantum} {Computers}}.
\newblock \emph{\bibinfo{journal}{arXiv:2203.13739 [quant-ph]}}
  (\bibinfo{year}{2022}).
\newblock \urlprefix\url{http://arxiv.org/abs/2203.13739}.
\newblock \bibinfo{note}{ArXiv: 2203.13739}.

\bibitem{mps_classically_efficient}
\bibinfo{author}{Vidal, G.}
\newblock \bibinfo{title}{Efficient classical simulation of slightly entangled
  quantum computations}.
\newblock \emph{\bibinfo{journal}{Phys. Rev. Lett.}}
  \textbf{\bibinfo{volume}{91}}, \bibinfo{pages}{147902}
  (\bibinfo{year}{2003}).
\newblock
  \urlprefix\url{https://link.aps.org/doi/10.1103/PhysRevLett.91.147902}.

\bibitem{isometric_peps}
\bibinfo{author}{Zaletel, M.~P.} \& \bibinfo{author}{Pollmann, F.}
\newblock \bibinfo{title}{Isometric tensor network states in two dimensions}
  (\bibinfo{year}{2019}).
\newblock \eprint{arXiv:1902.05100}.

\bibitem{haghshenas2019conversion}
\bibinfo{author}{Haghshenas, R.}, \bibinfo{author}{O'Rourke, M.~J.} \&
  \bibinfo{author}{Chan, G. K.-L.}
\newblock \bibinfo{title}{Conversion of projected entangled pair states into a
  canonical form}.
\newblock \emph{\bibinfo{journal}{Physical Review B}}
  \textbf{\bibinfo{volume}{100}}, \bibinfo{pages}{054404}
  (\bibinfo{year}{2019}).

\bibitem{smith2021crossing}
\bibinfo{author}{Smith, A.}, \bibinfo{author}{Jobst, B.},
  \bibinfo{author}{Green, A.~G.} \& \bibinfo{author}{Pollmann, F.}
\newblock \bibinfo{title}{Crossing a topological phase transition with a
  quantum computer}.
\newblock \emph{\bibinfo{journal}{Phys. Rev. Research}}
  \textbf{\bibinfo{volume}{4}}, \bibinfo{pages}{L022020}
  (\bibinfo{year}{2022}).
\newblock
  \urlprefix\url{https://link.aps.org/doi/10.1103/PhysRevResearch.4.L022020}.

\bibitem{barratt2021parallel}
\bibinfo{author}{Barratt, F.} \emph{et~al.}
\newblock \bibinfo{title}{Parallel quantum simulation of large systems on small
  {NISQ} computers}.
\newblock \emph{\bibinfo{journal}{npj Quantum Information}}
  \textbf{\bibinfo{volume}{7}}, \bibinfo{pages}{1--7} (\bibinfo{year}{2021}).
\newblock \urlprefix\url{https://www.nature.com/articles/s41534-021-00420-3}.
\newblock \bibinfo{note}{Number: 1 Publisher: Nature Publishing Group}.

\bibitem{Lin2021real}
\bibinfo{author}{Lin, S.-H.}, \bibinfo{author}{Dilip, R.},
  \bibinfo{author}{Green, A.~G.}, \bibinfo{author}{Smith, A.} \&
  \bibinfo{author}{Pollmann, F.}
\newblock \bibinfo{title}{Real- and imaginary-time evolution with compressed
  quantum circuits}.
\newblock \emph{\bibinfo{journal}{PRX Quantum}} \textbf{\bibinfo{volume}{2}},
  \bibinfo{pages}{010342} (\bibinfo{year}{2021}).
\newblock \urlprefix\url{https://link.aps.org/doi/10.1103/PRXQuantum.2.010342}.

\bibitem{foss2021entanglement}
\bibinfo{author}{Foss-Feig, M.} \emph{et~al.}
\newblock \bibinfo{title}{Entanglement from tensor networks on a trapped-ion
  quantum computer}.
\newblock \emph{\bibinfo{journal}{Phys. Rev. Lett.}}
  \textbf{\bibinfo{volume}{128}}, \bibinfo{pages}{150504}
  (\bibinfo{year}{2022}).
\newblock
  \urlprefix\url{https://link.aps.org/doi/10.1103/PhysRevLett.128.150504}.

\bibitem{hastings2007area}
\bibinfo{author}{Hastings, M.~B.}
\newblock \bibinfo{title}{An area law for one-dimensional quantum systems}.
\newblock \emph{\bibinfo{journal}{Journal of Statistical Mechanics: Theory and
  Experiment}} \textbf{\bibinfo{volume}{2007}}, \bibinfo{pages}{P08024--P08024}
  (\bibinfo{year}{2007}).
\newblock \urlprefix\url{https://doi.org/10.1088/1742-5468/2007/08/p08024}.
\newblock \bibinfo{note}{Publisher: IOP Publishing}.

\bibitem{hastings2007entropy}
\bibinfo{author}{Hastings, M.~B.}
\newblock \bibinfo{title}{Entropy and entanglement in quantum ground states}.
\newblock \emph{\bibinfo{journal}{Phys. Rev. B}} \textbf{\bibinfo{volume}{76}},
  \bibinfo{pages}{035114} (\bibinfo{year}{2007}).
\newblock \urlprefix\url{https://link.aps.org/doi/10.1103/PhysRevB.76.035114}.

\bibitem{reyes2021multi}
\bibinfo{author}{Reyes, J.~A.} \& \bibinfo{author}{Stoudenmire, E.~M.}
\newblock \bibinfo{title}{Multi-scale tensor network architecture for machine
  learning}.
\newblock \emph{\bibinfo{journal}{Machine Learning: Science and Technology}}
  \textbf{\bibinfo{volume}{2}}, \bibinfo{pages}{035036} (\bibinfo{year}{2021}).
\newblock \urlprefix\url{https://doi.org/10.1088/2632-2153/abffe8}.
\newblock \bibinfo{note}{Publisher: IOP Publishing}.

\bibitem{stoudenmire2018learning}
\bibinfo{author}{Stoudenmire, E.~M.}
\newblock \bibinfo{title}{Learning relevant features of data with multi-scale
  tensor networks}.
\newblock \emph{\bibinfo{journal}{Quantum Science and Technology}}
  \textbf{\bibinfo{volume}{3}}, \bibinfo{pages}{034003} (\bibinfo{year}{2018}).
\newblock \urlprefix\url{https://doi.org/10.1088/2058-9565/aaba1a}.
\newblock \bibinfo{note}{Publisher: IOP Publishing}.

\bibitem{mera_vidal}
\bibinfo{author}{Vidal, G.}
\newblock \bibinfo{title}{Class of quantum many-body states that can be
  efficiently simulated}.
\newblock \emph{\bibinfo{journal}{Phys. Rev. Lett.}}
  \textbf{\bibinfo{volume}{101}}, \bibinfo{pages}{110501}
  (\bibinfo{year}{2008}).
\newblock
  \urlprefix\url{https://link.aps.org/doi/10.1103/PhysRevLett.101.110501}.

\bibitem{cong2019quantum}
\bibinfo{author}{Cong, I.}, \bibinfo{author}{Choi, S.} \&
  \bibinfo{author}{Lukin, M.~D.}
\newblock \bibinfo{title}{Quantum convolutional neural networks}.
\newblock \emph{\bibinfo{journal}{Nature Physics}}
  \textbf{\bibinfo{volume}{15}}, \bibinfo{pages}{1273--1278}
  (\bibinfo{year}{2019}).
\newblock \urlprefix\url{https://www.nature.com/articles/s41567-019-0648-8}.
\newblock \bibinfo{note}{Number: 12 Publisher: Nature Publishing Group}.

\bibitem{martyn2020entanglement}
\bibinfo{author}{Martyn, J.}, \bibinfo{author}{Vidal, G.},
  \bibinfo{author}{Roberts, C.} \& \bibinfo{author}{Leichenauer, S.}
\newblock \bibinfo{title}{Entanglement and {Tensor} {Networks} for {Supervised}
  {Image} {Classification}}.
\newblock \emph{\bibinfo{journal}{arXiv:2007.06082 [quant-ph, stat]}}
  (\bibinfo{year}{2020}).
\newblock \urlprefix\url{http://arxiv.org/abs/2007.06082}.
\newblock \bibinfo{note}{ArXiv: 2007.06082}.

\bibitem{crowley2014quantum}
\bibinfo{author}{Crowley, P. J.~D.},
  \bibinfo{author}{\DJ{}uri\ifmmode~\acute{c}\else \'{c}\fi{}, T.},
  \bibinfo{author}{Vinci, W.}, \bibinfo{author}{Warburton, P.~A.} \&
  \bibinfo{author}{Green, A.~G.}
\newblock \bibinfo{title}{Quantum and classical dynamics in adiabatic
  computation}.
\newblock \emph{\bibinfo{journal}{Phys. Rev. A}} \textbf{\bibinfo{volume}{90}},
  \bibinfo{pages}{042317} (\bibinfo{year}{2014}).
\newblock \urlprefix\url{https://link.aps.org/doi/10.1103/PhysRevA.90.042317}.

\bibitem{barratt2021dissipative}
\bibinfo{author}{Barratt, F.} \emph{et~al.}
\newblock \bibinfo{title}{Dissipative failure of adiabatic quantum transport as
  a dynamical phase transition}.
\newblock \emph{\bibinfo{journal}{Phys. Rev. A}}
  \textbf{\bibinfo{volume}{103}}, \bibinfo{pages}{052427}
  (\bibinfo{year}{2021}).
\newblock \urlprefix\url{https://link.aps.org/doi/10.1103/PhysRevA.103.052427}.

\bibitem{barratt2022improvements}
\bibinfo{author}{Barratt, F.}, \bibinfo{author}{Dborin, J.} \&
  \bibinfo{author}{Wright, L.}
\newblock \bibinfo{title}{Improvements to gradient descent methods for quantum
  tensor network machine learning}.
\newblock \emph{\bibinfo{journal}{arXiv preprint arXiv:2203.03366}}
  (\bibinfo{year}{2022}).
\newblock \urlprefix\url{https://arxiv.org/abs/2203.03366}.

\bibitem{dilip_data_2022}
\bibinfo{author}{Dilip, R.}, \bibinfo{author}{Liu, Y.-J.},
  \bibinfo{author}{Smith, A.} \& \bibinfo{author}{Pollmann, F.}
\newblock \bibinfo{title}{Data compression for quantum machine learning}.
\newblock \emph{\bibinfo{journal}{arXiv:2204.11170 [cond-mat,
  physics:quant-ph]}}  (\bibinfo{year}{2022}).
\newblock \urlprefix\url{http://arxiv.org/abs/2204.11170}.
\newblock \bibinfo{note}{ArXiv: 2204.11170}.

\bibitem{schuld2019machine}
\bibinfo{author}{Schuld, M.}
\newblock \bibinfo{title}{Machine learning in quantum spaces}
  (\bibinfo{year}{2019}).

\bibitem{lloyd2020quantum}
\bibinfo{author}{Lloyd, S.}, \bibinfo{author}{Schuld, M.},
  \bibinfo{author}{Ijaz, A.}, \bibinfo{author}{Izaac, J.} \&
  \bibinfo{author}{Killoran, N.}
\newblock \bibinfo{title}{Quantum embeddings for machine learning}.
\newblock \emph{\bibinfo{journal}{arXiv preprint arXiv:2001.03622}}
  (\bibinfo{year}{2020}).
\newblock \urlprefix\url{https://arxiv.org/abs/2001.03622}.

\bibitem{larose_robust_2020}
\bibinfo{author}{LaRose, R.} \& \bibinfo{author}{Coyle, B.}
\newblock \bibinfo{title}{Robust data encodings for quantum classifiers}.
\newblock \emph{\bibinfo{journal}{Phys. Rev. A}}
  \textbf{\bibinfo{volume}{102}}, \bibinfo{pages}{032420}
  (\bibinfo{year}{2020}).
\newblock \urlprefix\url{https://link.aps.org/doi/10.1103/PhysRevA.102.032420}.

\bibitem{SteveWhiteDMRG}
\bibinfo{author}{White, S.~R.}
\newblock \bibinfo{title}{Density matrix formulation for quantum
  renormalization groups}.
\newblock \emph{\bibinfo{journal}{Phys. Rev. Lett.}}
  \textbf{\bibinfo{volume}{69}}, \bibinfo{pages}{2863--2866}
  (\bibinfo{year}{1992}).
\newblock \urlprefix\url{https://link.aps.org/doi/10.1103/PhysRevLett.69.2863}.

\bibitem{huang2021power}
\bibinfo{author}{Huang, H.-Y.} \emph{et~al.}
\newblock \bibinfo{title}{Power of data in quantum machine learning}.
\newblock \emph{\bibinfo{journal}{Nature Communications}}
  \textbf{\bibinfo{volume}{12}}, \bibinfo{pages}{2631} (\bibinfo{year}{2021}).
\newblock \urlprefix\url{https://www.nature.com/articles/s41467-021-22539-9}.
\newblock \bibinfo{note}{Number: 1 Publisher: Nature Publishing Group}.

\bibitem{huang2021information}
\bibinfo{author}{Huang, H.-Y.}, \bibinfo{author}{Kueng, R.} \&
  \bibinfo{author}{Preskill, J.}
\newblock \bibinfo{title}{Information-theoretic bounds on quantum advantage in
  machine learning}.
\newblock \emph{\bibinfo{journal}{Phys. Rev. Lett.}}
  \textbf{\bibinfo{volume}{126}}, \bibinfo{pages}{190505}
  (\bibinfo{year}{2021}).
\newblock
  \urlprefix\url{https://link.aps.org/doi/10.1103/PhysRevLett.126.190505}.

\bibitem{cotler2021revisiting}
\bibinfo{author}{Cotler, J.}, \bibinfo{author}{Huang, H.-Y.} \&
  \bibinfo{author}{McClean, J.~R.}
\newblock \bibinfo{title}{Revisiting dequantization and quantum advantage in
  learning tasks}.
\newblock \emph{\bibinfo{journal}{arXiv:2112.00811 [quant-ph]}}
  (\bibinfo{year}{2021}).
\newblock \urlprefix\url{http://arxiv.org/abs/2112.00811}.
\newblock \bibinfo{note}{ArXiv: 2112.00811}.

\bibitem{harrow_quantum_2009}
\bibinfo{author}{Harrow, A.~W.}, \bibinfo{author}{Hassidim, A.} \&
  \bibinfo{author}{Lloyd, S.}
\newblock \bibinfo{title}{Quantum algorithm for linear systems of equations}.
\newblock \emph{\bibinfo{journal}{Phys. Rev. Lett.}}
  \textbf{\bibinfo{volume}{103}}, \bibinfo{pages}{150502}
  (\bibinfo{year}{2009}).
\newblock
  \urlprefix\url{https://link.aps.org/doi/10.1103/PhysRevLett.103.150502}.

\bibitem{dervovic_quantum_2018}
\bibinfo{author}{Dervovic, D.} \emph{et~al.}
\newblock \bibinfo{title}{Quantum linear systems algorithms: a primer}.
\newblock \emph{\bibinfo{journal}{arXiv:1802.08227}}  (\bibinfo{year}{2018}).
\newblock \urlprefix\url{http://arxiv.org/abs/1802.08227}.

\bibitem{song_tree_2019}
\bibinfo{author}{Cheng, S.}, \bibinfo{author}{Wang, L.},
  \bibinfo{author}{Xiang, T.} \& \bibinfo{author}{Zhang, P.}
\newblock \bibinfo{title}{Tree tensor networks for generative modeling}.
\newblock \emph{\bibinfo{journal}{Phys. Rev. B}} \textbf{\bibinfo{volume}{99}},
  \bibinfo{pages}{155131} (\bibinfo{year}{2019}).
\newblock \urlprefix\url{https://link.aps.org/doi/10.1103/PhysRevB.99.155131}.

\bibitem{wall_tree-tensor-network_2021}
\bibinfo{author}{Wall, M.~L.} \& \bibinfo{author}{D'Aguanno, G.}
\newblock \bibinfo{title}{Tree-tensor-network classifiers for machine learning:
  From quantum inspired to quantum assisted}.
\newblock \emph{\bibinfo{journal}{Phys. Rev. A}}
  \textbf{\bibinfo{volume}{104}}, \bibinfo{pages}{042408}
  (\bibinfo{year}{2021}).
\newblock \urlprefix\url{https://link.aps.org/doi/10.1103/PhysRevA.104.042408}.

\end{thebibliography}

\appendix

\section{Tree tensor classifier}\label{app:TreeTensor}

Here we elaborate on the tree tensor network (TTN) version of the classifier. Tree tensor networks were originally proposed in the study of many-body systems where entanglement is accumulated in a hierarchical fashion\cite{shi_classical_2006}. They have also been used as machine learning models\cite{stoudenmire2018learning, song_tree_2019, reyes2021multi} and are similarly suitable for implementation on a quantum computer as matrix product states\cite{huggins2019towards, hierarchical_quantum_classifiers, wall_tree-tensor-network_2021}. The primary alteration in our case is in the different decomposition of the amplitude encoded image state, $\ket{\psi}$, shown in Fig.~\ref{fig:MPSEncoding} of the main text, and the resulting form of the batched classifier. We consider a decomposition {\it via} singular value decompositions (SVDs) into a tree structure with $N$ leaves (describing physical indices), $\{L^{\sigma_i}\}_{i=1}^N$, $N_B$ layers of branches, $\{B^{[j, \gamma]}\}$, where $j=1, \dots, N_B$ labels the branch layer and $\gamma$ labels each branch within the layer and finally a top trunk tensor, $T$. The SVD is performed up the tree from leaves to trunk, and as a result each tensor is an isometry. We describe this in Eqn.~(\ref{eqn:svd_tree_tensor}) showing auxiliary indices.

\begin{multline}
\label{eqn:svd_tree_tensor}
    \ket{\psi} \underset{\text{SVD}}{\longrightarrow} \sum_{\{a, b, \dots, t\}}\sum_{\{\sigma\}} \prod_{\Theta=0}^{\sfrac{(\eta-2)}{2}} T_{t_{2\Theta+1}t_{2\Theta+2}}\times\\
    \times \prod_{\eta=0}^{\sfrac{(\zeta-2)}{2}} B^{[N_B, \eta], t_{\eta}}_{s_{2\eta+1}, s_{2\eta+2}} \times \cdots \times \prod_{\alpha=0}^{\sfrac{(\beta-2)}{2}} B^{[2, \alpha], c_{\alpha}}_{b_{2\alpha+1}, b_{2\alpha+2}} \times \\
    \times \prod_{n=0}^{\sfrac{(N-2)}{2}} B^{[1, n], b_n}_{a_{2n+1}, a_{2n+2}} L^{a_{2n+1}}_{\sigma_{2n+1}}L^{a_{2n+2}}_{\sigma_{2n+2}}\ket{\sigma_{2n+1}\sigma_{2n+2}},
\end{multline}

\begin{figure}[h]
\includegraphics[width=0.85\columnwidth, height=0.9\columnwidth]{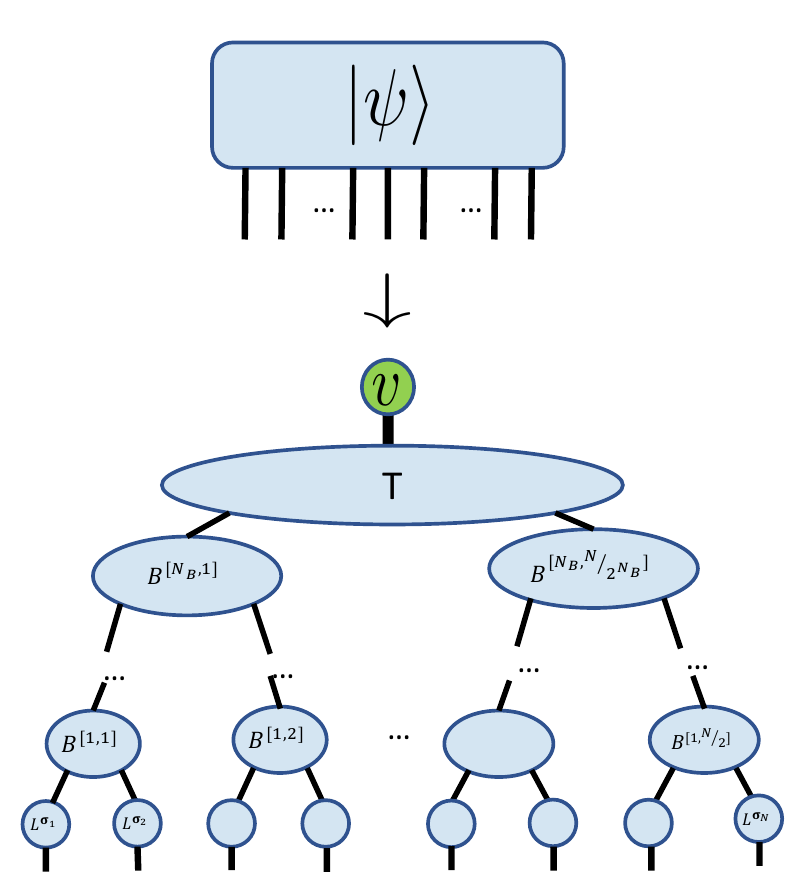}
\caption{{\bf Tree Tensor Network (TTN) encoding.} \newline
Starting from the same amplitude encoded state, $\ket{\psi}$ as in Fig.~\ref{fig:MPSEncoding}(a), 
a sequence of singular value decompositions\cite{mps_classically_efficient} transform the vector into a tree tensor form, with the isometries directed from leaves to trunk via a series of branches. Again, $v_\delta$ is a set of amplitudes for some $N$-spin reference state which we conventionally take to be $\ket{0}^{\otimes N}$. 
}
\label{fig:TTNEncoding}
\end{figure}

\subsection{Tree Tensor Encoding of the Sum State}
\label{app:SumStateTree}

Here we describe the analogous construction of the prototypical sum state in tree tensor form. The procedure is qualitatively the same as for the MPS, but we expound upon it for completeness.

For a set of three TTNs characterised by tensors:
\begin{align*}
    \{ \{L_{\sigma_n, 1}\}_{n=1}^{N},\{B^{[j, \gamma]}_1\}_{j=1}^{N_B}, \{T_1\}\}, \\
    \{ \{L_{\sigma_n, 2}\}_{n=1}^{N},\{B^{[j, \gamma]}_2\}_{j=1}^{N_B}, \{T_2\}\}, \\
    \{ \{L_{\sigma_n,3}\}_{n=1}^{N},\{B^{[j, \gamma]}_3\}_{j=1}^{N_B}, \{T_3\}\},
\end{align*}
the tree representation of the sum state is formed by first combining the tensors into one larger tensor as follows:\\

\noindent
i. Leaves:
\begin{align*}
 & \mathcal{L}_{\sigma_n, l_1\otimes l_2 \otimes l_3}\\
=& \left(L^{1}_{\sigma_n\; 1}, \dots, L^{D_1}_{\sigma_n\; 1}, 
L^{1}_{\sigma_n\; 2}, \dots, L^{D_1}_{\sigma_n\; 2}, L^{1}_{\sigma_n\: 3}, \dots, L^{D_1}_{\sigma_m\;3}\right)\\
=& \left(\underline{L}_{\sigma_n, 1},\underline{L}_{\sigma_n, 2},\underline{L}_{\sigma_n, 3}\right)
\end{align*}
ii. Branches:
\begin{equation*}
\mathcal{B}^{[j, \gamma]}
=
\left(
\begin{array}{ccc}
\underline{\underline{B}}^{[j, \gamma]}_{1} & 0 & 0\\
0 & \underline{\underline{B}}^{[j, \gamma]}_{2}  & 0 \\
0 & 0 & \underline{\underline{B}}^{[j, \gamma]}_{3} 
\end{array}
\right)
\end{equation*}
iii. Trunk:
\begin{equation*}
\mathcal{T}=
\left(
\begin{array}{ccc}
\underline{\underline{T}}_{1} & 0 & 0\\
0 & \underline{\underline{T}}_{2}  & 0 \\
0 & 0 & \underline{\underline{T}}_{3} 
\end{array}
\right),
\end{equation*}
where $\underline{M}$ is a vector and $\underline{\underline{M}}$ is a matrix. Auxiliary-space tensor indices have been suppressed in the second and third cases for clarity.

\begin{figure}
          \includegraphics[width = 0.9\columnwidth, height=1.2\columnwidth]{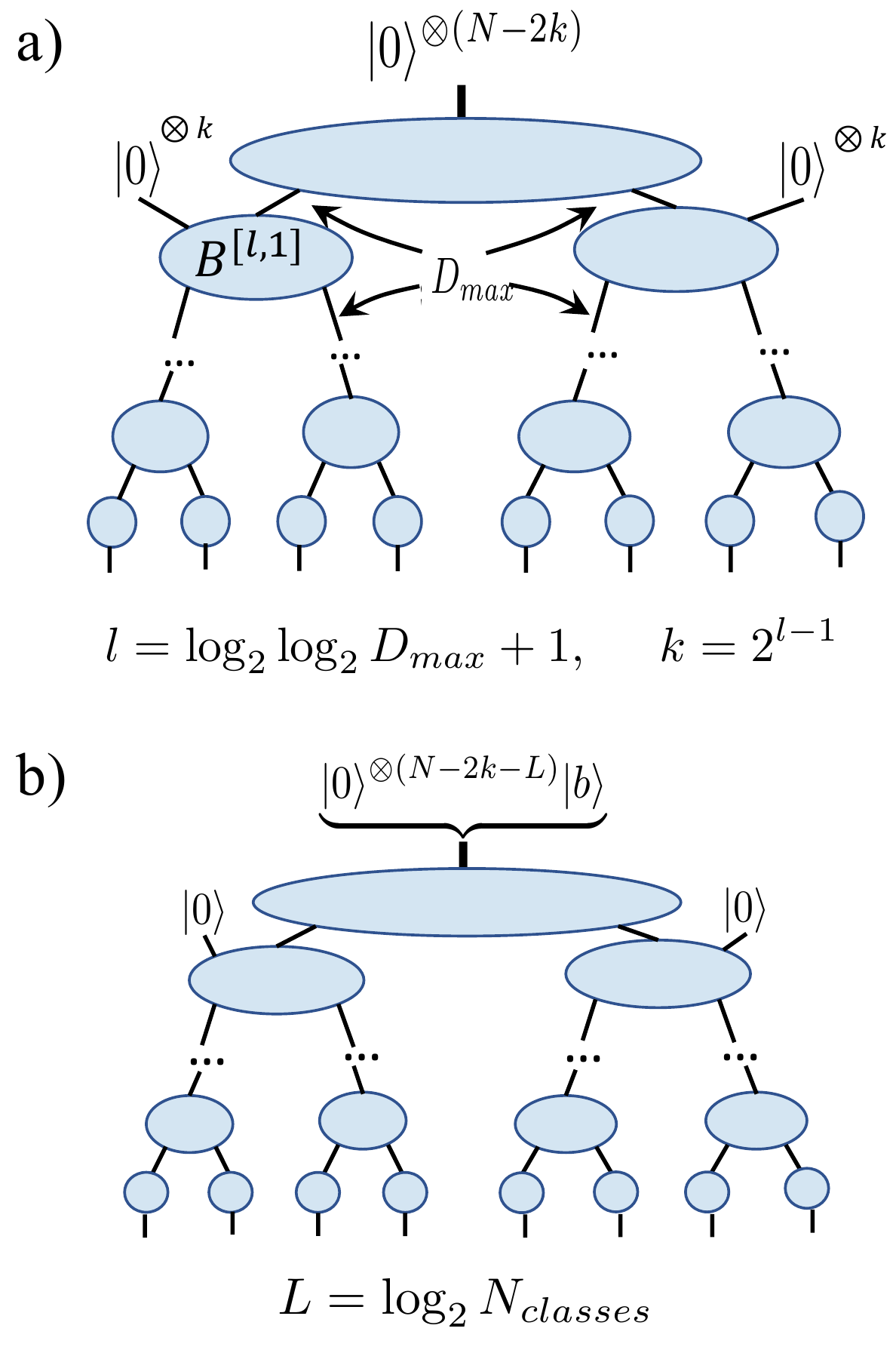}
    \caption{
    {\bf Compression and labelling of images in tree tensor network form.} \newline
    a) {\it Compression of TTN:} The amplitude encoding can be compressed by truncating a sequence of singular value decompositions on the auxiliary space from leaves to trunk\cite{schollwock2011density,orus_review,Perez-Garcia2006} to some maximum value $D_{\max}$. This implies a corresponding reshaping of the tensors and the reference state as shown. If a truncation occurs after the $l^{th}$ branch, this corresponds to a projection of $2^{l-1}$ qubits onto a reference state. \newline
    b) {\it Labelling:} As with the MPS, the use of different reference states provides a straightforward way to label wavefunction encodings of images in different classes. The final $\log_2 N_{\text{classes}}$ reference states are used for a binary label, the first $\log_2 (N_{\text{pixels}}/N_{\text{classes}})$ remaining as $\ket{0}$. It is convenient to reshape the tensors as indicated so that the label is entirely on the final tensor. This reshaping facilitates multi-state classification on a quantum circuit with a single unitary.}
    \label{fig:TTNCompression}
\end{figure}

\subsection{Quantum Circuit Tree Tensors} \label{app:tree_tensor_circuit}

\begin{figure}[h]
\centering
\includegraphics[width=0.7\columnwidth]{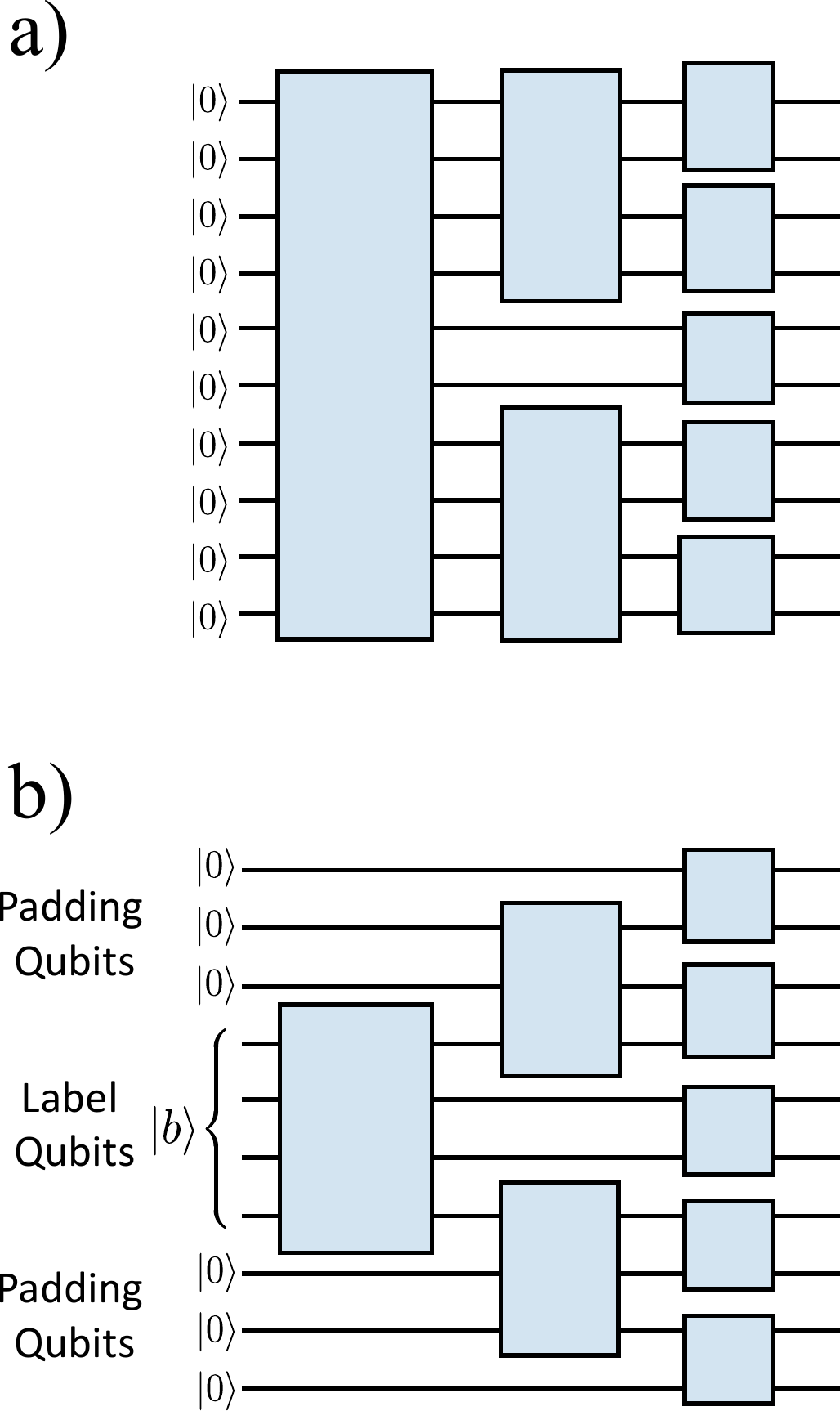}
\caption{{\bf Circuit Representation of TTN-encoded images.} A TNN in orthogonal form can be translated directly to a quantum circuit, similarly to a MPS. a) shows the translation of the full quantum state $\ket{\psi}$ and is equivalent to Fig.~\ref{fig:TTNCompression}b) shows a truncation to bond order $D_{\text{max}}=4$ and how reference qubits are used to label the classes. 
}
\label{fig:CircuitRepresentationTree}
\end{figure}

In Fig.~\ref{fig:CircuitRepresentationTree} we give a quantum circuit representation of the tree tensor network, analogous to Fig.~\ref{fig:CircuitRepresentation} in the main text for the MPS. For simplicity, we depict a version with $10$ qubits, and note that trunk of the TTN to which this circuit corresponds would have three downward legs. One could perform a final SVD on two of these legs, leaving an extra tensor, but we omit this for symmetry reasons.

One notable feature to observe is the TTN decomposes into fewer unitaries than the MPS as a result of the faster (doubly exponential) growth of tensor bond dimension from the leaves to the trunk. While this hierarchical structure may be able to capture certain correlations in the data more effectively than the linear decomposition of the MPS, it may be less desirable for an actual quantum circuit implementation. This is because the circuit requirements of the trunk tensor (centre tensor in the MPS) will dominate the gate complexity of any decomposition into native gates, for example the control NOT gate, and so it is the primary target for truncation. However, since the TTN decomposition results in fewer unitaries, it may be that truncating the trunk tensor is more detrimental to the classifier accuracy than truncating the centre tensor in the MPS. Another way of viewing this is observing that the SVD procedure into a tensor network decomposition (and truncation) is itself a form of (approximate) quantum circuit compilation. Note than both in the TTN and the centre-gauge MPS, the largest unitary will be the same size (a $10$ qubit unitary for MNIST for example). Therefore, larger gains in quantum circuit performance may be achievable by investigating left or right canonical MPS (instead of the mixed canonical we use here) whose unitary forms will differ. However, we leave further analysis of the tradeoff between these architectures to future work.

\section{Tuning Feature Extraction}\label{app:tuning_feature_extraction}
In the main text we demonstrate a deterministic scheme to initialise a tensor network feature extractor.
This was done by sequentially adding batches of data encoded as tensor networks, compressing and then orthogonalising using a combination of SVDs and a polar decomposition of the orthogonality centre.
Several variables exist that control the amount of compression at different points of the procedure.
These include $D_{\text{encode}}$, the bond order of the tensor network encoded images; $D_{\text{batch}}$, the bond order of the batch added encoded images and $D_{\text{final}}$, the bond order of the finalised feature extractor.
\\
In the following sections, we tune these parameters for the different datasets and show a redundancy in the computational resources required without loss of performance.

\subsection{Fashion MNIST}
\begin{figure*}[ht!]
\centering
\includegraphics[width = \linewidth]{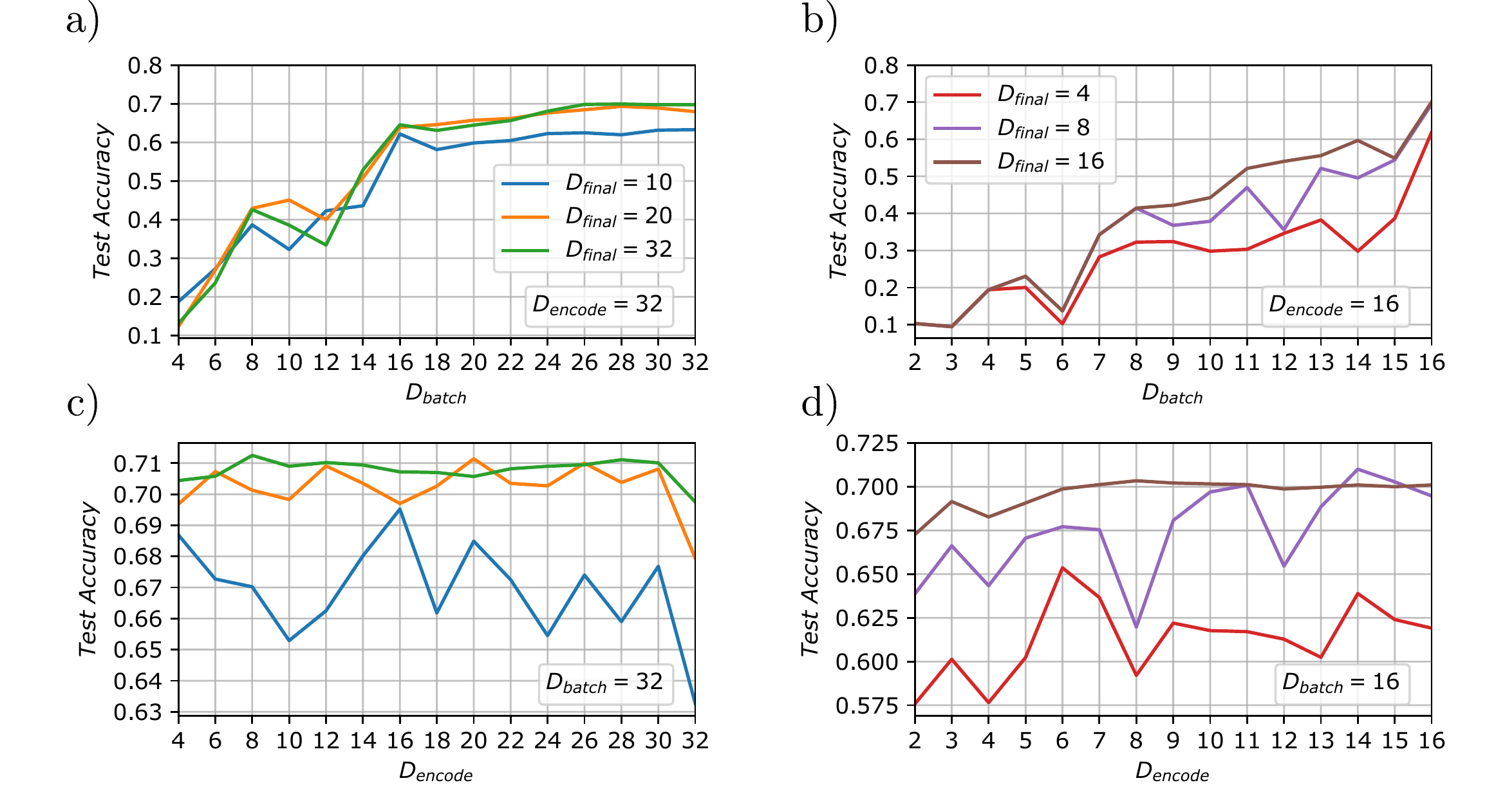}
\caption{ 
\textbf{Test Accuracy for Different Encoding and Batching Bond Orders on Fashion MNIST}.\\
Test accuracy of orthogonal classifier for various $D_{\text{batch}}$ (a, b) and $D_{\text{encode}}$ (c, d). {\it Left column} (a, c) shows results for the MPO decomposition, and {\it right column} (b, d) shows results for TTO decomposition. In all cases the hyperparameter which is not varied (e.g. $D_{\text{batch}}$ in (c, d)) is set to maximal, $32$ in the case of the MPO, $16$ for the TTO. Here we see MPO classifiers with $D_{\text{final}} = 20$, $D_{\text{final}} = 32$ perform quite similarly, suggesting a redundancy in information for the latter case.
}
\label{fig:fashion_acc_vs_d_batch_d_encode}
\end{figure*}

Fig.~\ref{fig:fashion_acc_vs_d_batch_d_encode} shows the performance for varying $D_{\text{encode}}$ and $D_{\text{batch}}$ for the MPO and TTO feature extractors respectively with a fixed $D_{\text{final}}$.
We can see the performance of the feature extractors is independent on the value of $D_{\text{encode}}$, but highly dependent on the value of $D_{\text{batch}}$.
We believe that the strong dependence upon $D_{batch}$ reflects a necessity to explore a larger state space in order to reach the final classifier which may be representable at a lower bond order.

\subsection{MNIST}
\begin{figure*}[ht!]
\centering
\includegraphics[width = 0.9\linewidth]{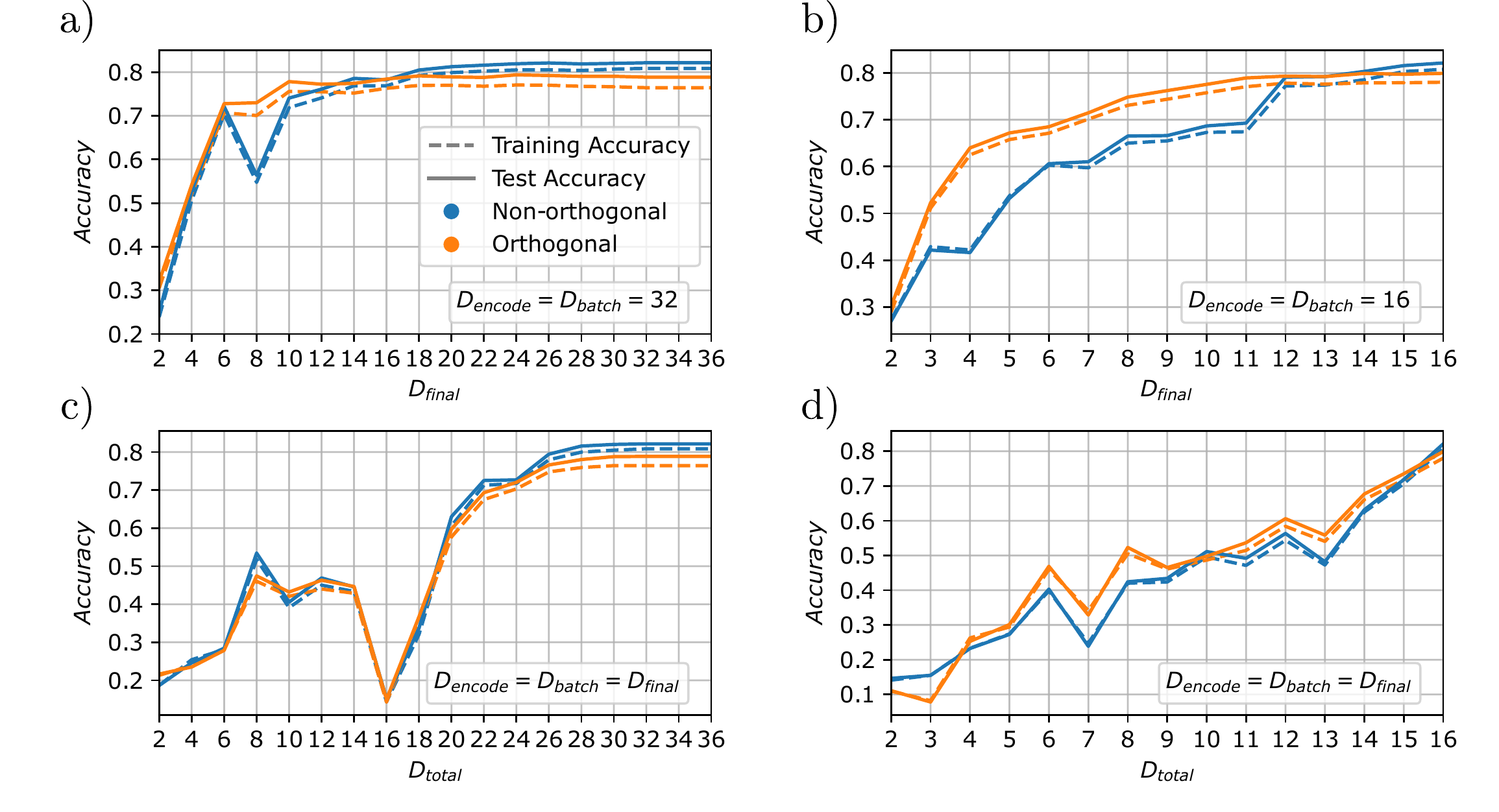}
\caption{ 
\textbf{Classification Accuracy {\it versus} Bond Order for MNIST.}\\
MNIST training and test accuracy of the deterministically-constructed tensor network classifiers for different bond orders, using the (a, c) MPO and (b, d) TTO decompositions. For the MPO (TTO), $54210$ ($54080$) training images were used to construct the classifier, whilst all $10000$ test images were used to evaluate performance.
{\it Top row} (a, b) shows accuracy as a function of $D_{\text{final}}$ with $D_{\text{encode}}$ and $D_{\text{batch}}$ set to be maximal. {\it Bottom row} (c, d) shows $D_{\text{encode}}$ set equal to $D_{\text{batch}}$ and $D_{\text{final}}$ with all three varied simultaneously. We also show classifier performance with and without orthogonalisation (polar decomposition on the final MPO tensor, or trunk tensor of TTO) which is required for implementation on a quantum device. Analogous to Fig.~\ref{fig:fashion_accuracy_vs_d_total} in the main text.
}
\label{fig:mnist_accuracy_vs_d_total}
\end{figure*}

\begin{figure*}[ht!]
\centering
 \includegraphics[width = 0.9\linewidth]{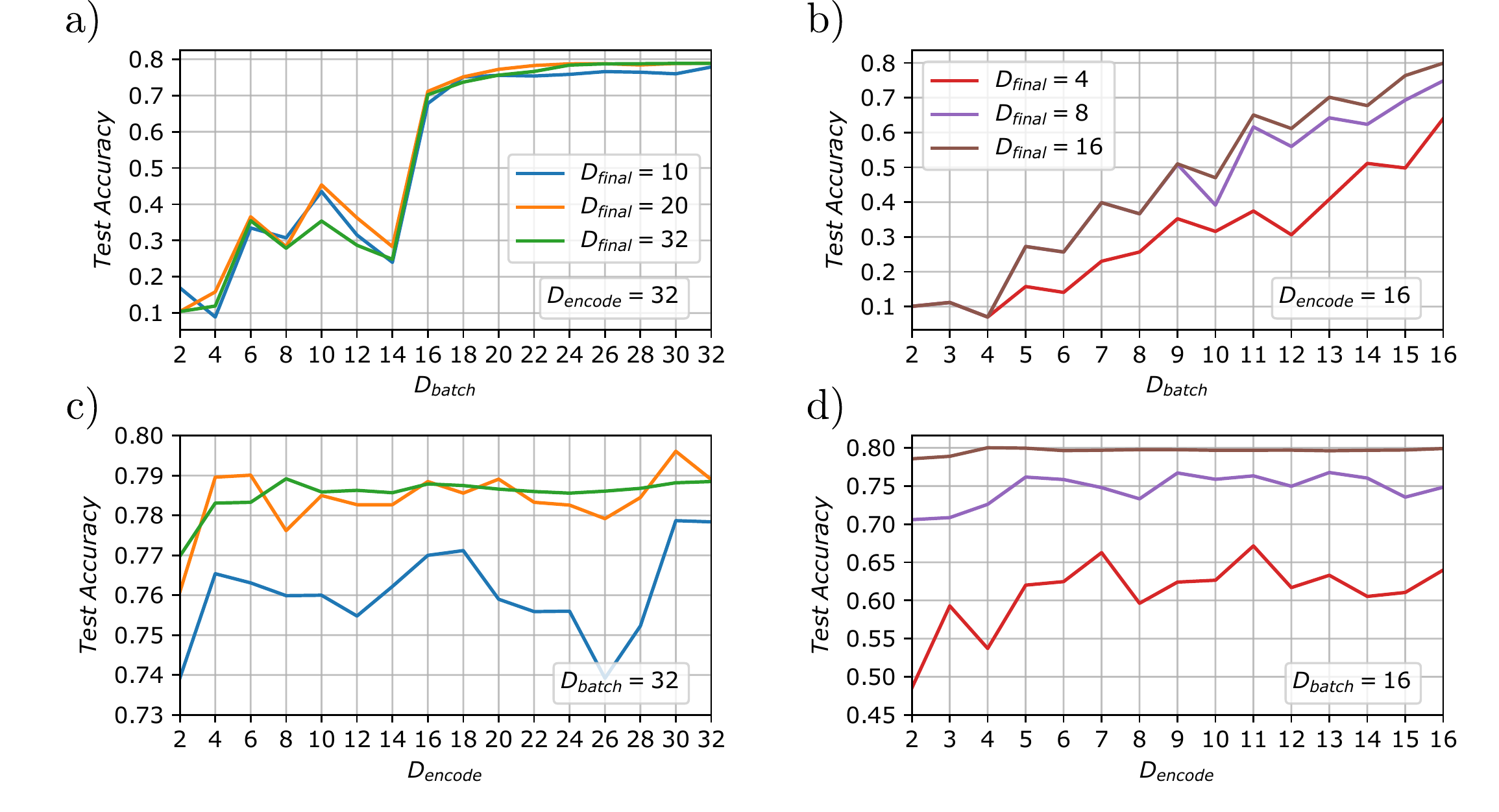}
\caption{ 
\textbf{Test Accuracy for Different Encoding and Batching Bond Orders on MNIST}.\\
MNIST test accuracy of orthogonal classifier for various $D_{\text{batch}}$ (a, b) and $D_{\text{encode}}$ (c, d). {\it Left column} (a, c) shows results for the MPO decomposition, and {\it right column} (b, d) shows results for TTO decomposition. In all cases the hyperparameter that is not varied (e.g. $D_{\text{batch}}$ in (c, d)) is set to maximal, $32$ in the case of the MPO, $16$ for the TTO. Here we see MPO classifiers with $D_{\text{final}} = 20$, $D_{\text{final}} = 32$ perform quite similarly, suggesting a redundancy in information for the latter case.}
\label{fig:mnist_acc_vs_d_batch_d_encode}
\end{figure*}

Fig.~\ref{fig:mnist_accuracy_vs_d_total} shows the MPO and TTO feature extractor performances for varying $D_{\text{final}}$ and $D_{\text{final}} = D_{\text{encode}} = D_{\text{batch}}$, whilst Fig.~\ref{fig:mnist_acc_vs_d_batch_d_encode} varies $D_{\text{encode}}$ and $D_{\text{batch}}$ for fixed $D_{\text{final}}$ for the MNIST dataset.
The results show similar trends to those obtained with the fashion MNIST dataset, plateauing at a higher accuracy due to the simpler nature of the dataset

\section{Quantum Stacking} \label{app:quantum_stacking}
In this section we present the results of quantum stacking for the MNIST dataset in Figs.~\ref{fig:mnist_stacking_mps_hierarchy} and \ref{fig:mnist_stacking_mps_hierarchy}b,  and visualise the effect of stacking on predictions in Fig.~\ref{fig:confusion_matrices}.

\vspace{0.1in}

{\it Test Accuracy:} The dependence of test accuracy upon the number of layers of hierarchical stacking, Fig.~\ref{fig:mnist_stacking_mps_hierarchy}a, and with the number of copies of the uploaded data in a tensor network stacking, Fig.~\ref{fig:mnist_stacking_mps_hierarchy}b, show similar trends to those found for the fashion MNIST dataset. The main difference is the higher ultimate performance found for this data set. 

\vspace{0.1in}
{\it Visualising the effects of stacking:} A confusion matrix is constructed to measure the output of the stacked circuit compared to the various bitstring labels;
\begin{equation}
{\cal C}_{ij}=
\left|\langle{\sigma_i}|^{\otimes M}
V ({\bm 1}^{M-1} \otimes |b_j\rangle 
\langle b_j \ket{\sigma_i}^{\otimes M} \right|^2,
\label{eq:confusion}
\end{equation}
where the image data has been uploaded $M$ times and the label space sum-state image $|\sigma_i \rangle = \bra{ 0}^{\otimes (N-L)} U |\Sigma_i \rangle $ is used as a proxy for images in the class $i$. In the case of a single upload of the data and no stacking, Eq.(\ref{eq:confusion}) gives the fidelity between the bitstring and the label-space image. Fig.~\ref{fig:confusion_matrices} shows a greyscale representation of the elements of the confusion matrices with different degrees of stacking. 
We have permuted the class labels to reveal these matrices in their most diagonal form  (We do so by minimizing the cost function $|{\cal C} - W|$, where ${\cal C}$ 
is the unstacked confusion matrix and $W$ is a weight matrix with values that decrease monotonically away from the leading diagonal).
As the degree of stacking increases, the matrices become increasingly diagonal implying an increase in orthogonality between the different label states.
As the degree of stacking tends to infinity, we expect the confusion matrices to become completely diagonal.

\begin{widetext}

\begin{figure}[ht!]
\includegraphics[width = 0.9\textwidth]{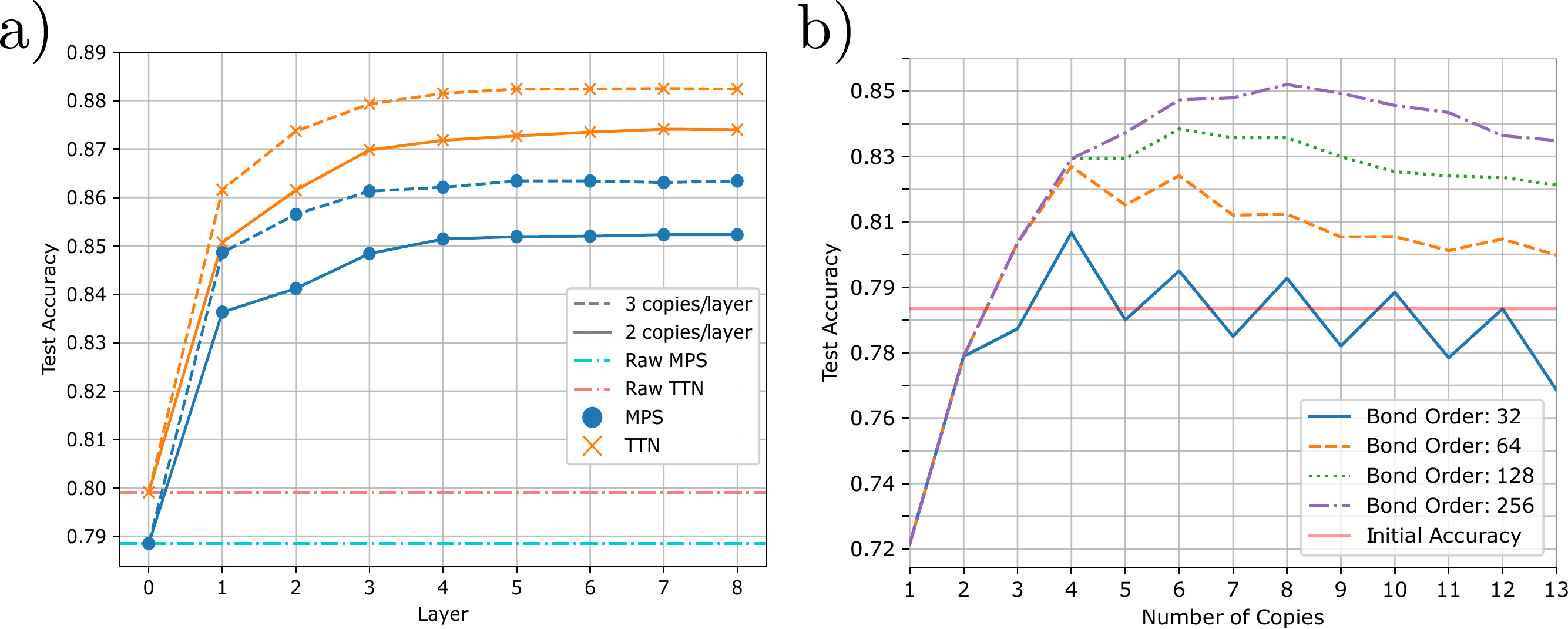}
\caption{\textbf{Stacking results for MNIST.}\\
a) {\it Hierarchical Quantum Stacking.}
Hierarchical stacking resulting from sequential solutions of Eq.~\ref{eq:InitialiseV} for a small number of copies in each $V_i$. Layer $0$ corresponds to the raw accuracy for the original MPO/TTO classifier for the MNIST dataset, analogous to Fig.~\ref{fig:hierarchical_stacking_fashion} in the main text. Again, we see an improvement with increasing layers, and number of copies per layer.\\ b) {\it Tensor Network Quantum Stacking.} 
The test accuracy of quantum stacking as a function of the number of copies of the data uploaded and the bond order of the tensor network approximations to the stacking unitary $V$. Increasing the number of copies leads to an increase in performance- regulated by bond order. $V$ is constructed from the batch addition of the copied training image predictions encoded as MPS as described in the main text.
This figure is analogous to Fig.\ref{fig:mps_stacking} in the main text for fashion MNIST}
\label{fig:mnist_stacking_mps_hierarchy}
\end{figure}

\begin{figure}[ht!]
    \includegraphics[width = 0.95\textwidth]{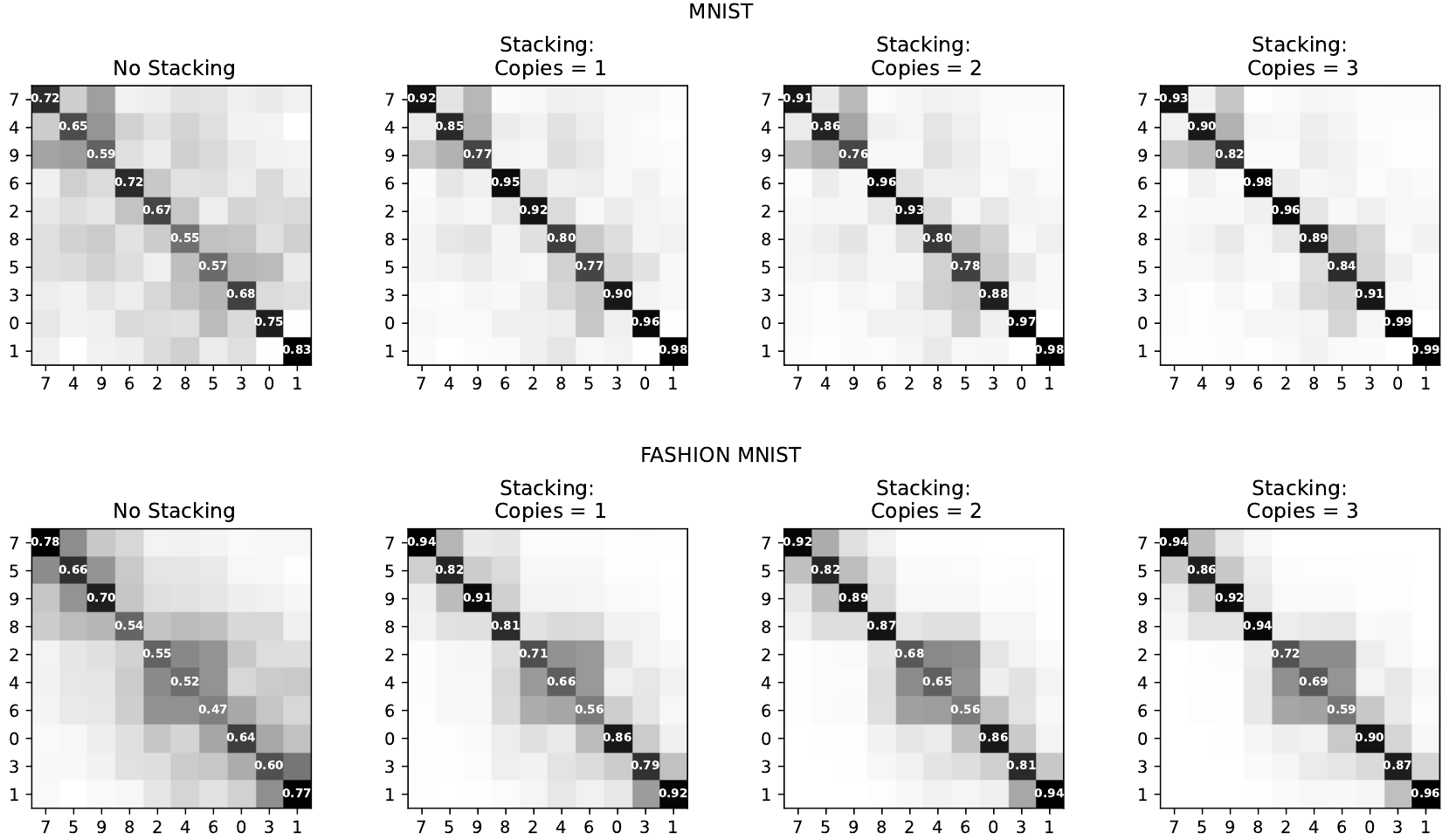}
    \caption{\textbf{Evolution of confusion matrices with refinement by quantum stacking}. The figures shows a grey-scale representation of the confusion matrices given by Eq.(\ref{eq:confusion}) and how they evolve with the degree of quantum stacking. The columns and rows of the confusion matrix have been permuted to find its most diagonal form. Rearranging the confusion matrix in this way allows insight into any clustering/commonalities between labels. Increasing the number of copies in the stacking procedure leads to more diagonal confusion matrices, implying an increase contrast between different labels. This increase in contrast is particularly noticeable for MNIST, which matches the higher increase in performance from quantum stacking when compared to fashion MNIST.}
    \label{fig:confusion_matrices}
\end{figure}

\end{widetext}
\end{document}